\documentclass[aps,showpacs,preprintnumbers,amsmath,amssymb]{revtex4}
\UseRawInputEncoding
\usepackage{dcolumn}
\usepackage[dvips]{epsfig}

\usepackage{float}
\usepackage{graphicx,epsfig}
\usepackage{graphicx}
\usepackage{epstopdf}
\usepackage{dcolumn}
\usepackage{bm}
\usepackage{gensymb}
\usepackage{footnote}
\usepackage{booktabs}

\begin{document}
\baselineskip=0.8 cm
\title{Image of a regular phantom compact object and its luminosity under spherical accretions}

\author{Xin Qin$^{1}$
Songbai Chen$^{1,2}$\footnote{Corresponding author: csb3752@hunnu.edu.cn},
Jiliang Jing$^{1,2}$ \footnote{jljing@hunnu.edu.cn}}
\affiliation{ $ ^1$ Department of Physics, Key Laboratory of Low Dimensional Quantum Structures
and Quantum Control of Ministry of Education, Synergetic Innovation Center for Quantum Effects and Applications, Hunan
Normal University,  Changsha, Hunan 410081, People's Republic of China
\\
$ ^2$Center for Gravitation and Cosmology, College of Physical Science and Technology, Yangzhou University, Yangzhou 225009, People's Republic of China}

\begin{abstract}
\baselineskip=0.6 cm
\begin{center}
{\bf Abstract}
\end{center}

We have studied the shadow and its luminosity for a static and regular phantom compact object under the static spherical accretion and the infalling spherical accretion, respectively. Comparing with the usual Schwarzschild black hole, the presence of phantom hair yields the larger black hole shadow and the darker image. In both spherical accretion models, with the increase of phantom parameter, the maximum luminosity occurred at photon ring and the brightness of the central region in the shadow decrease, but in the region far from the shadow, the luminosity of image slightly increases. The image of a phantom wormhole and its luminosity have similar behaviors.  Moreover, as the phantom charge parameter $\alpha$ increases up to the critical value at where the compact object changes from black hole to wormhole, there exists a jump for the specific intensity, which also appears in the slowly rotating case. This implies that the phantom hair is imprinted on both the shadow radius and the intensity of the electromagnetic flux radiation around compacted object.
\end{abstract}

 \maketitle
\newpage
\section{Introduction}

The first image of the supermassive black hole in the center of M87 galaxy released by Event Horizon Telescope
Collaboration \cite{EHT1,EHT2,EHT3} is one of the most exciting events in observing black holes. The main ingredients
in the image are the black hole shadow and the luminosity distribution of radiating gas around the black hole, which could help us to further understand black hole and the matter accretion process around black hole. The black hole shadow, a dark interior in the image, which is believed to encode the fingerprint of the geometry near the black hole, has been applied to constrain black hole parameters \cite{eht1,nkerr} and extra dimension size \cite{extr1,extr2}, and to probe some fundamental physics issues including dark matter \cite{tomoch,dark1,dark2,dark3,Boson,dark4}, the equivalence principle \cite{epb} and the no-hair theorem \cite{sw,swo,astro,chaotic}.

In general, the real astrophysical black holes in the galaxies are surrounded by accretion matters. It is expectable that the distribution of accretion matters should play an important role in black hole image in this case. The
first image of a black hole with an emitting accretion thin disk was obtained in \cite{grr0}, which shows that there are primary and secondary images of the thin accretion disk appeared outside black hole shadow. The image for a Kerr black hole with Keplerian accretion disk has been simulated in Refs. \cite{grr,short,BKD}.  Vincent \textit{et al} \cite{gyoto} has investigated the images of a thin infinite accretion disk and Ion torus around compact objects. The effects of the interaction between black hole and accretion matter on the shadow were studied in \cite{chensa,cunha}, which indicates that the shadow becomes more prolate as the accretion mass grows and the non-integrable of photon motion arising from the interaction leads to self-similar fractal structures appeared in black hole shadow. Moreover, for a Schwarzschild black hole with thin and thick accretion disks, the recent investigation \cite{Gralla} shows that the photon ring together with the lensing ring make significant contribution to the black hole shadow and the corresponding observed flux. Although the real accretion flows are generically not spherically symmetric, the simplified spherical model could capture key features of accretion appeared in the usual general-relativistic magnetohydrodynamics models \cite{EHT2}. With this spirit, the spherical accretion model has been applied to analyze the image of a Schwarzschild black hole \cite{termed,NJG}. It is shown that the size of the observed shadow can serve as a signature of the spacetime geometry since the location of the shadow edge is independent of the inner radius at which the accreting gas stops radiating and is unrelated to the details of accretion process. The effects of spherical accretion on the shadow for a four-dimensional Gauss-Bonnet black hole has been studied in \cite{xz}, which indicates that the Gauss-Bonnet constant enhances the observed specific intensity of the black hole image. For a black hole with dark matter hair \cite{sau}, the intensity of the electromagnetic flux radiation in the spherical accretion depends on the dark matter model, and the dark matter distribution plays an important role in the shadow radius and the intensity of the electromagnetic flux radiation. Moreover, the effects of quintessence dark energy on image of black hole on the intensity of flux radiation is also investigated in the spherical accretion \cite{xz1}.

Phantom scalar field is another theoretical model of dark energy with the super-negative equation of state $\omega <-1$, which leads to the violation of the null energy condition, the big rip of the universe \cite{phantom}. However, the phantom dark energy can not excluded by the current precise observational data \cite{phantom1}.
Some black hole solutions describing gravity coupled to phantom scalar fields or phantom Maxwell fields have been found and the corresponding geometric structure and thermodynamic properties are also studied in \cite{BKZ,phantom3,phantom4,phantom5,phantom6,phantom7,phantom8,phantom9,phantom10,phantom11}. Extending the regular phantom black hole solution to the case of slow rotation is studied in \cite{phantom12}. The strong gravitational lensing of such a kind of black holes with phantom hair has been investigated in \cite{lensing1,lensing2,lensing3,lensing4}. Moreover, it is also interesting to consider the phantom type of black holes by looking into their possible theoretical origin. It might be achieved by introducing higher order gravitational operators. For example, the phantom type of black holes are obtained in the high-order asymptotically safe gravity \cite{phbhole1} and in the quadratic gravity \cite{phbhole2}. Given that the previous images store the fingerprint of black holes,
it is natural to ask whether we can obtain the information about phantom field by the image of a black hole together with its luminosity distribution under the spherical accretions. In this paper, we will study the image of a regular and static compacted object with phantom scalar hair \cite{BKZ} under the static spherical accretion and the infalling spherical accretion, respectively. In the static spherical accretion the radiating gas is assumed to be at rest, but in the infalling one the gas is supposed to fall freely towards black hole along the radial direction. With these two simple accretion models, we want to probe the effect of phantom field on the compacted object image and on the light intensity measured by a distant observer. We also generalize it to a slowly rotating case.

The paper is organized as follows: In Sec.II, we investigate the effect of phantom parameter on the image of such a regular compacted object and its luminosity under the static spherical accretion model and the infalling spherical accretion model, respectively. In Sec.III,  we study the light intensity of a regular slowly rotating phantom compacted object under the static spherical accretion model and the infalling spherical accretion model, respectively.  Finally, we end the paper with a summary.

\section{Image of a regular and static phantom compact object under spherical accretions}

Let us now review in brief a regular and static solution with phantom scalar hair obtained by Bronnikov \textit{et al} \cite{BKZ}. The action with phantom scalar field $\Phi$ in Einstein gravity can be expressed as \cite{BKZ}
\begin{equation}\label{Action}
S=\int\sqrt{-g}d^{4}x\left[R-g^{\mu\nu}\partial_{\mu}\Phi\partial_{\nu}\Phi+2V\left(\Phi\right)\right],
\end{equation}
where $\Phi$ is the phantom scalar field with negative kinetic energy and $V$ is the potential for the scalar field. Here, we take $8\pi G=c=1$.
From the action (\ref{Action}), Bronnikov et al. obtained a regular and static solution with phantom scalar hair, whose metric form can be written as
\begin{equation}\label{metric}
ds^2=-F(r^\prime)dt^2+\frac{1}{F(r^\prime)}d{r^\prime}^2+\left({r^\prime}^2+\alpha^2\right)
\left(d\theta^2+\sin^2{\theta}d\phi^2\right),
\end{equation}
with
\begin{equation}
F(r^\prime)=1-\frac{3M}{\alpha}\left[\left(\frac{\pi}{2}-\arctan\frac{r'}{\alpha}\right)\left(1+\frac{r'^2}{\alpha^2}\right)-\frac{r'}{\alpha}\right].
\end{equation}
The scalar field is $\Phi\equiv\sqrt{2}\psi=\sqrt{2}\arctan\frac{r^\prime}{\alpha}$, and the corresponding potential has a form $V=\frac{3M}{\alpha^3}[(\frac{\pi}{2}-\psi)(3-2\cos^2{\psi})-3\sin\psi\cos\psi]$.
The quantity $M$ is the usual Schwarzschild-like mass and the positive constant $\alpha$ is
related to the charge of phantom scalar field. The event horizon is located at position where $F(r^\prime)=0$. It is shown that
the radius of the event horizon $r_h$ is in the range $0< r_h\leq2M $ as $0\leq\alpha<3\pi M/2$ and becomes $r_h=0$ as $\alpha=3\pi M/2$.  When $\alpha>3\pi M/2$ , one can find that the value of $r_h$ is negative, which means that
there does not exist any event horizon and then a throat appears as in wormholes.
Thus, the phantom hair brings about richer properties for the spacetime (\ref{metric}). Moreover, the metric \eqref{metric} can reduce to Schwarzschild one as $\alpha=0$ because the phantom scalar field $\Phi$ becomes a constant $\frac{\sqrt{2}\pi}{2}$ and the potential disappears, which means that the action (\ref{Action}) turns to the usual action without any material field  in this limit.

To adopt conveniently the technique of studying the spherical accretions in Refs. \cite{NJG,xz,sau,xz1},  for the compact object with phantom scalar hair (\ref{metric}),  we can introduce a new coordinate $r^2={r^\prime}^2+\alpha^2$ and turn the metric form (\ref{metric}) into
 \begin{equation}\label{newmetric}
ds^2=-F(r)dt^2+\frac{r^2}{F(r)\left(r^2-\alpha^2\right)}dr^2+r^2\left(d\theta^2+\sin^2{\theta}d\phi^2\right),
\end{equation}
with
\begin{eqnarray}\label{specificFC}
F(r)=1-\frac{3M}{\alpha}\left[\left(\frac{\pi}{2}-\arctan\frac{\sqrt{r^2-\alpha^2}}{\alpha}\right)
\frac{r^2}{\alpha^2}-\frac{\sqrt{r^2-\alpha^2}}{\alpha}\right].
\end{eqnarray}
\begin{figure}
\centering
  \includegraphics[width=5cm ]{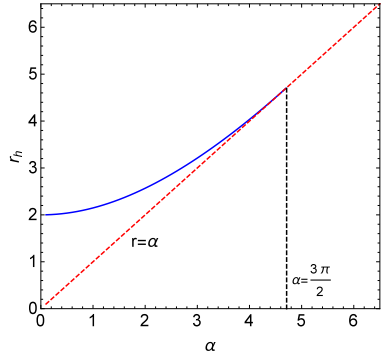}
\caption{The change of the event horizon radius $r_{h}$ with phantom parameter $\alpha$ for the regular and static phantom black hole. Here we set $M=1$.}
\label{as00}
\end{figure}
The sphere part in the new metric form (\ref{newmetric}) has the same geometry as that in the usual static metric, which ensures that the forms of the specific emissivity and of the power per-unit volume in Refs. \cite{NJG,xz,sau,xz1} for the radiating gas in spherical accretion can be applied in the case of a regular  compact object with phantom hair.
It is obvious that for the spacetime (\ref{newmetric}) the physicals range of $r$ is $r>\alpha$. When $0<\alpha<\frac{3\pi{M}}{2}$, we find that the radius of event horizon $r_h$ is in the range $2M<r_h<\frac{3\pi{M}}{2}$ and $r_h>\alpha$ is always satisfied, which is shown in Fig.(\ref{as00}). However, as $\alpha>\frac{3\pi{M}}{2}$, we find that there exists no the event horizon and the surface $r=\alpha$ owns the behavior of a throat in wormholes,  which means that the spacetime (\ref{newmetric}) does not describe the geometry of a black hole as in the metric (\ref{metric}). In other words, with the increase of $\alpha$, the spacetime (\ref{newmetric}) changes gradually from a black hole to a wormhole. Thus, studying image of a compact object (\ref{newmetric}) is helpful to understand the difference in observable information of black hole and wormhole.
Moreover, as $\alpha\rightarrow0$, we can find that the metric (\ref{newmetric}) can also reduces to the Schwarzschild metric.

\begin{figure}
\centering
\includegraphics[width=6cm ]{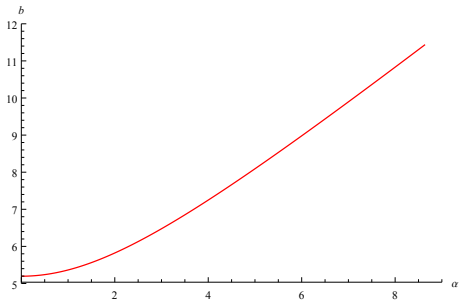}
\caption{The change of the impact parameter $b_{ph}$ with phantom parameter $\alpha$ for the regular and static phantom black hole. Here we set $M=1$.}
\label{as0}
\end{figure}

\begin{figure}
\includegraphics[width=5.1cm ]{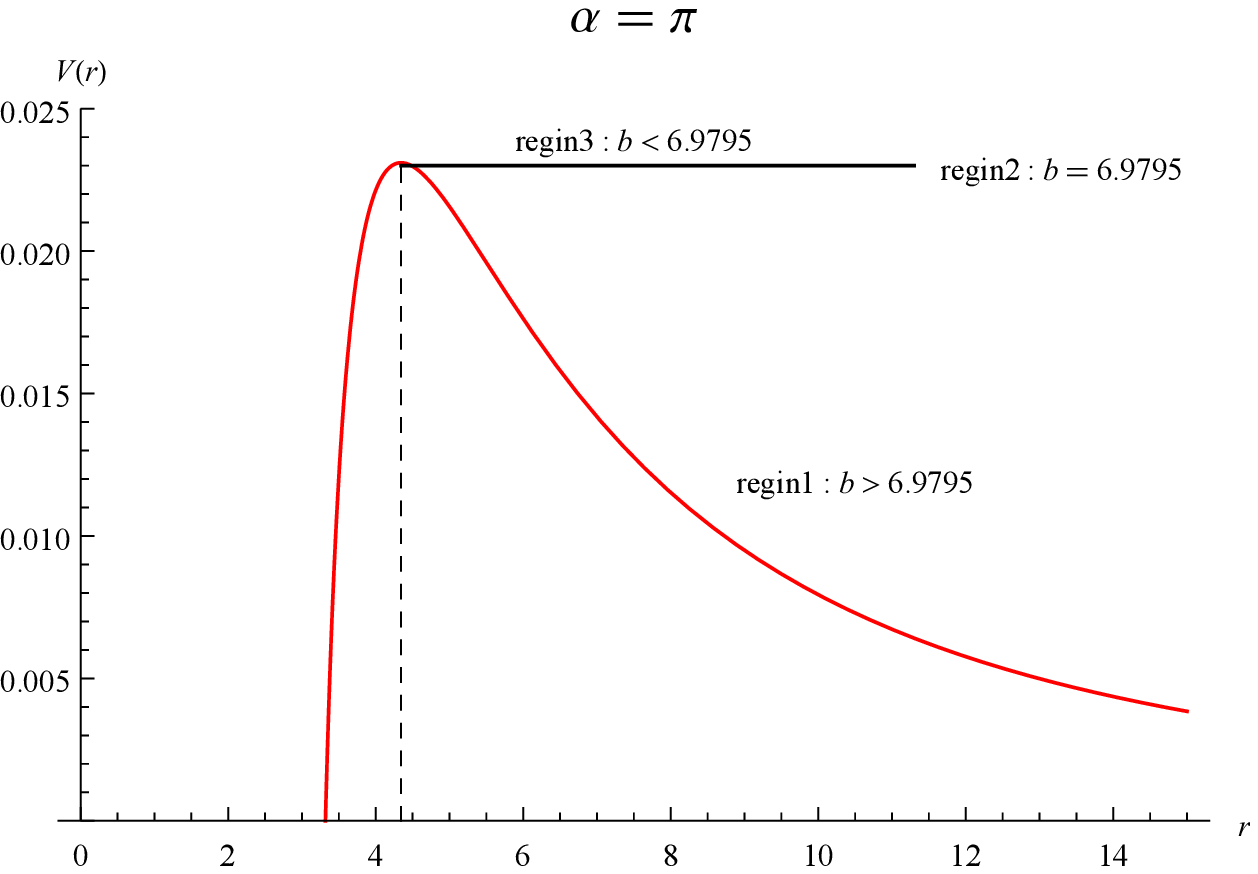}\quad\quad\includegraphics[width=5.1cm ]{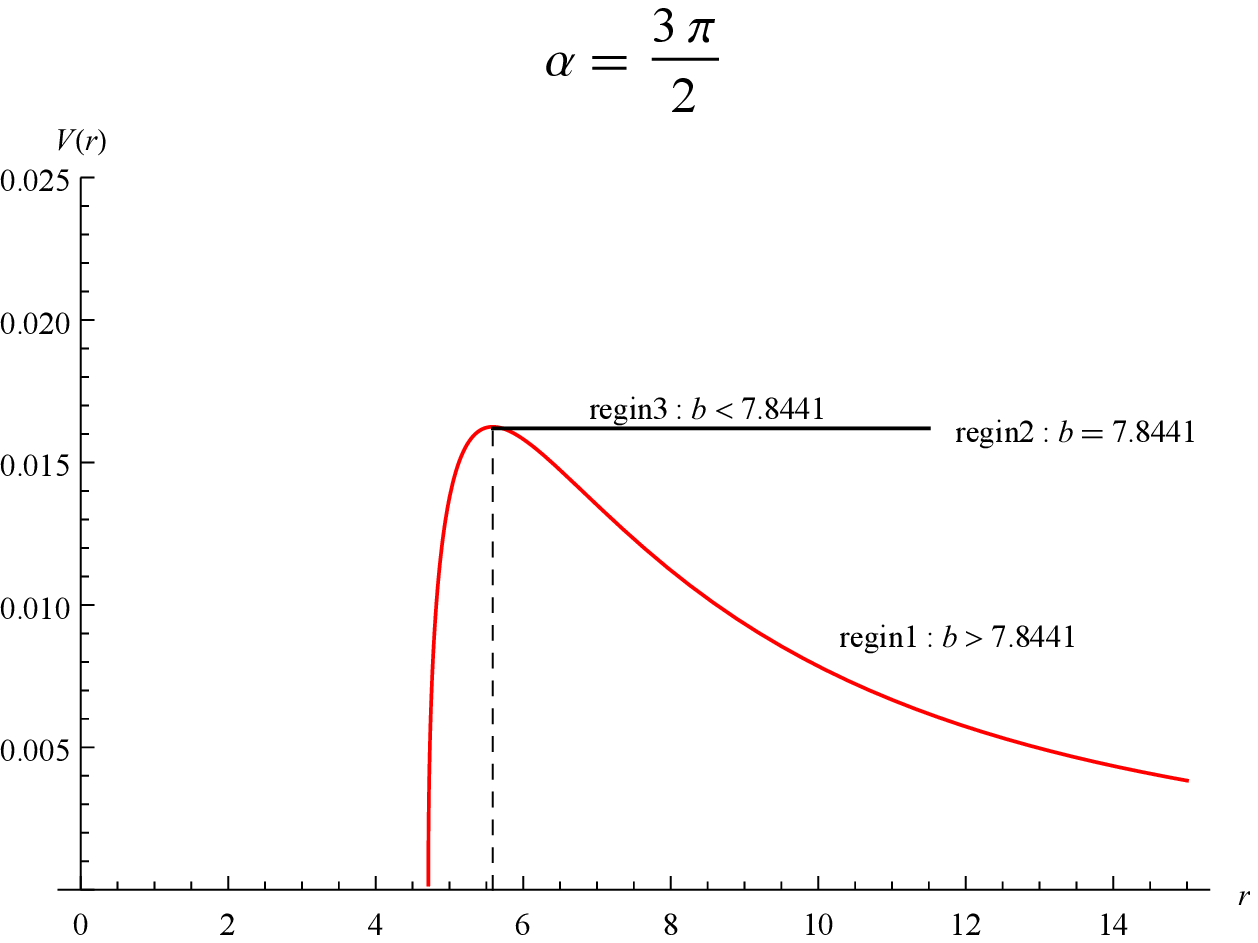}\quad\quad\includegraphics[width=5.1cm ]{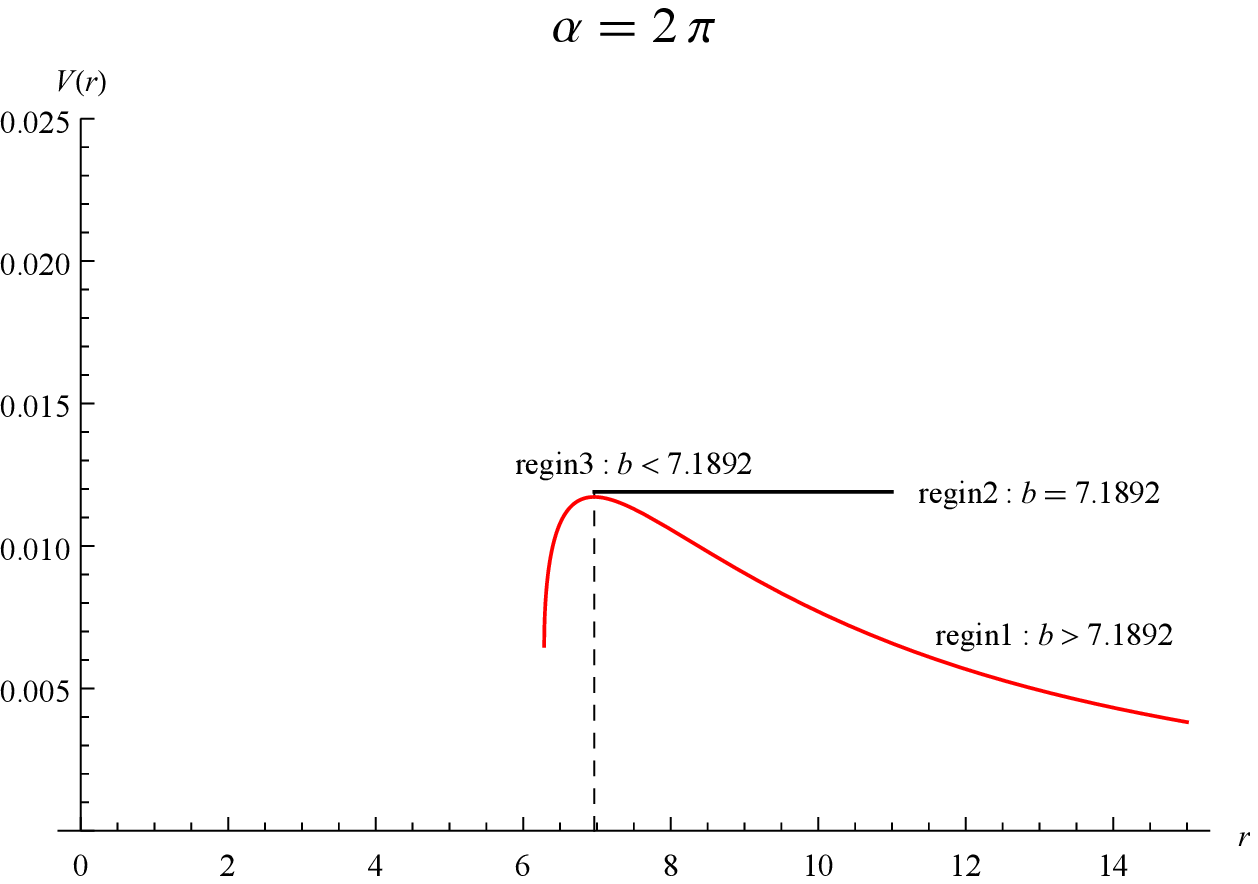}
\caption{The change of the effective potential with the polar coordinate $r$ for the fixed $\alpha$. The figures from left to right correspond to $\alpha=\pi$, $\alpha=\frac{3\pi}{2}$ and $\alpha=2\pi$, respectively. Here we set $M=1$}
\label{as01}
\end{figure}

Let us now to study firstly the null geodesics around a regular and static compact object with phantom scalar hair (\ref{newmetric}).
With the Lagrangian density for photon
\begin{equation}\label{Lagrangian}
\mathcal{L}=\frac{1}{2}g_{\mu\nu}\dot{x}^\mu\dot{x}^\nu=\frac{1}{2}\left(-F(r)\dot{t}^2+
\frac{r^2}{F(r)\left(r^2-\alpha^2\right)}\dot{r}^2+r^2\dot{\theta}^2+r^2\sin^{2}\theta\dot{\phi}^2\right),
\end{equation}
and the Lagrangian equation
 \begin{equation}\label{E-L equation}
\frac{d}{d\lambda}\left(\frac{\partial{\mathcal{L}}}{\partial\dot{x}^\mu}\right)=\frac{\partial{\mathcal{L}}}{\partial{x}^\mu},
\end{equation}
one can obtain the geodesics of photon moving in the spacetime of regular and static compact object with phantom scalar hair (\ref{newmetric}). Here $\lambda$ is an affine parameter and the dot denotes the derivative with respect to the parameter $\lambda$. It is obvious that there are two conserved quantities
since the metric functions in spacetime \eqref{newmetric} are not functions of coordinates $t$ and $\phi$.
These two conserved quantities corresponds to photon's
energy $E$ and angular momentum $L$, respectively. Their forms are
\begin{eqnarray}\label{energy and angular}
&&E\equiv-p_t=-\frac{\partial{L}}{\partial{\dot{t}}}=F(r)\frac{dt}{d\lambda},  \\ \nonumber
&&L\equiv p_{\phi}=\frac{\partial{L}}{\partial{\dot{\phi}}}=r^2\sin^2{\theta}\frac{d\phi}{d\lambda}.
\end{eqnarray}
For a static and spherical spacetime, we focus only on the photon moving on the equatorial plane, i.e.,  the constrained condition $\theta=\frac{\pi}{2}$ and $\dot{\theta}=0$ is satisfied for the photon. With the redefined affine parameter $\lambda\rightarrow\frac{\lambda}{|L|}$ and a impact parameter $b=\frac{|L|}{E}$, one can obtain the four-velocity of photon in the spacetime \eqref{newmetric}
 \begin{eqnarray}\label{energy}
&&\dot{t}=\frac{1}{bF(r)},\\
\label{angel}
&&\dot{\phi}=\pm\frac{1}{r^2},\\
\label{radial}
&& \frac{r^2}{\left(r^2-\alpha^2\right)}\dot{r}^2+\frac{F(r)}{r^2}=\frac{1}{b^2}.
\end{eqnarray}
The signs $+$ and $-$ represent photon moving in counterclockwise and clockwise along its azimuth, respectively.
From Eq.\eqref{radial}, we can obtain the photon sphere satisfy $\dot{r}=0$ and $\ddot{r}$=0, i.e.,
\begin{eqnarray}\label{V}
&&\frac{F(r)}{r^2}=\frac{1}{b^2},\nonumber\\
&&F'(r)r-2F(r)=0.
\end{eqnarray}
The largest root of the equation (\ref{V}) is the radius of the photon sphere $r_{ph}$, which has a form
\begin{eqnarray}\label{phs}
r_{ph}=\sqrt{9M^2+\alpha^2}.
\end{eqnarray}
The corresponding impact parameter $b_{ph}$ of photon moving along photon sphere is
\begin{eqnarray}\label{bhs}
b_{ph}=\sqrt{\frac{2\alpha^3}{2\alpha-3M\pi+6M\arctan\frac{3M}{\alpha}}}.
\end{eqnarray}
As $\alpha$ increases, it is easy to find that the photon sphere radius $r_{ph}$ and the impact parameter $b_{ph}$ increase, which is also shown in Fig.(\ref{as0}).
In Fig.(\ref{as01}), we plot the effective potential for $\alpha=\pi$, $\alpha=\frac{3\pi}{2}$ and $\alpha=2\pi$, respectively. Obviously, the effective potential reaches its maximum value at the photon sphere. The feature of effective potential dominates the motion of photon in the background spacetime.
For a photon moving in the radially inward  direction, if its impact parameter $b>b_{ph}$, it will encounter the potential barrier and be reflected back in the outward direction.  If its impact parameter $b=b_{ph}$, the photon will asymptotically approach the photon sphere and  revolve around the compact object many times. However, the orbit of photon at photon sphere is not stable. Thus,
photon will fall into the compact object or escape to the spatial infinity as long as its motion is slightly disturbed. When the impact parameter $b<b_{ph}$, the motion of photon in a black hole background is different from that in a wormhole background. In the case of a black hole (i.e., $0<\alpha<\frac{3\pi{M}}{2}$), the photon directly crosses the barrier and eventually falls into the black hole, while in the case of a wormhole $\alpha>\frac{3\pi{M}}{2}$, the photon will reach the throat and be reflected back to the original space-time and travel to infinity to reach the observer.
From Fig.(\ref{as01}), we find that the peak value of effective potential decreases with the increase of $\alpha$. However, the radial coordinate value at the position of the peak and the corresponding impact parameter $b_{ph}$ increase with  $\alpha$. These features of the effective potential of photon will affect the compact object shadow and the luminosity distribution around compact object.
Especially, for a spherically symmetrical black hole, its size of shadow is related directly to the size of the impact parameter. Therefore, the radius of the shadow for the distant static observer becomes larger in the case with the larger $\alpha$ value.

\subsection{Image of a regular phantom compact object with the static spherical accretion}

We are in position to study the image of a regular phantom compact object with a spherical accretion, which can be regarded to be optically thin.  For convenience, we here consider only two simple relativistic spherical accretion models, i.e., the static spherical accretion and the infalling spherical accretion. The radiating gas is assumed to be rest in the static spherical accretion and to infall freely into the compact object along the radial direction in the infalling one.

\begin{figure}
\centering
\includegraphics[width=6cm ]{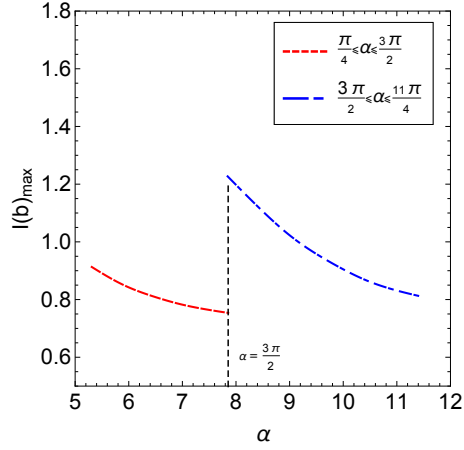}
\caption{The change of the the peak value of the specific intensity $I(b)$ with phantom parameter $\alpha$. Here we set $M=1$.}
\label{lo1}
\end{figure}

For the static spherical accretion, the observed specific intensity at the observed photon
frequency $\nu_{obs}$ for the pixel in the observer's image (measured usually in $erg s^{-1} cm^{-2} str^{-1} Hz^{-1}$) can be obtained by integrating the specific emissivity along the photon path $\gamma$, which
can be expressed as \cite{expression1,expression2}
\begin{equation}\label{intensity}
I(\nu_{obs})=\int_{\gamma}g^{3}j(\nu_{e})dl_{prop}.
\end{equation}
where $\nu_{e}$ is the photon frequency at the emitter and $g$ is the redshift factor , $dl_{prop}$ and $j(\nu_{e})$ are the infinitesimal proper length and the emissivity per-unit volume measured in the rest-frame of the emitter, respectively. For the  spacetime of a regular phantom compact object \eqref{newmetric}, the redshift factor has a form
\begin{equation}\label{redshift factor}
g=\frac{\nu_{obs}}{\nu_e}=F(r)^{\frac{1}{2}}.
\end{equation}
As in Refs.\cite{NJG,xz,sau,xz1}, one can assume that the emission is monochromatic with emitter's rest frame frequency $\nu_{\star}$ and the emission has a $1/r^2$ radial profile. And then the specific emissivity $j(\nu_{e})$ can be written as
\begin{equation}\label{emissivity}
j(\nu_{e})=\frac{\delta(\nu_e-\nu_\star)}{r^2},
\end{equation}
where $\delta$ is the delta function. In the spacetime \eqref{newmetric}, the proper length is
\begin{eqnarray}\label{proper length}
dl_{prop}=\sqrt{\frac{r^2}{F(r)\left(r^2-\alpha^2\right)}dr^2+r^2d\phi^2}.
\end{eqnarray}
With these above equations, the specific intensity (\ref{intensity}) measured by the observer at infinity can be further rewritten as
\begin{equation}\label{the radiation intensity}
I(\nu_{obs})=\int_{\gamma}\frac{F(r)^{\frac{3}{2}}}{r^2}\sqrt{\frac{r^2}{F(r)\left(r^2-\alpha^2\right)}
+r^2\frac{d\phi^2}{dr^2}}dr,
\end{equation}
with
\begin{equation}\label{89}
\frac{dr}{d\phi}=\pm{\sqrt{\left(r^2-\alpha^2\right)}}\sqrt{\frac{r^2}{b^2}-F(r)}.\nonumber
\end{equation}
\begin{figure}
\includegraphics[width=6cm]{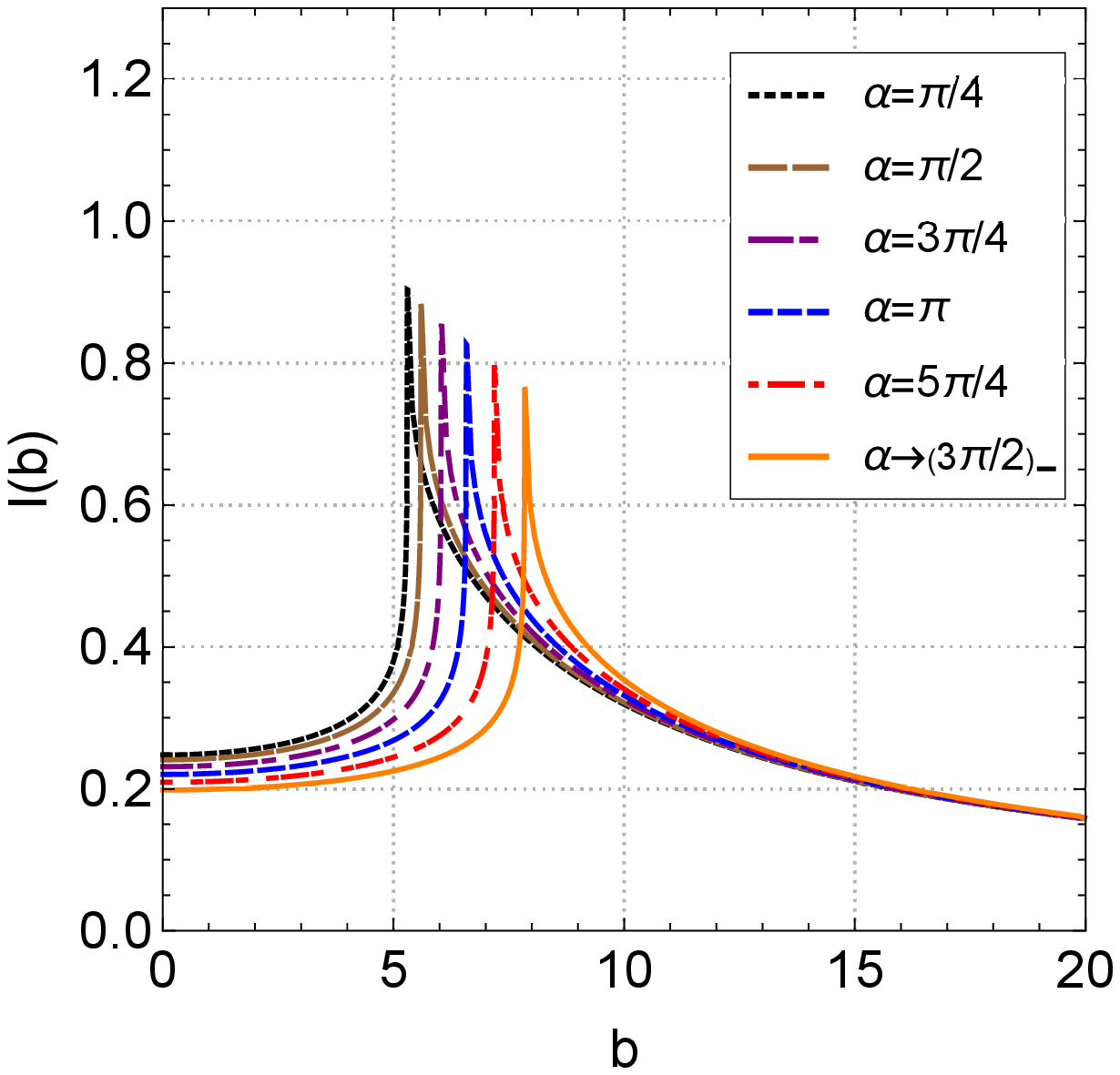}\quad\includegraphics[width=6cm]{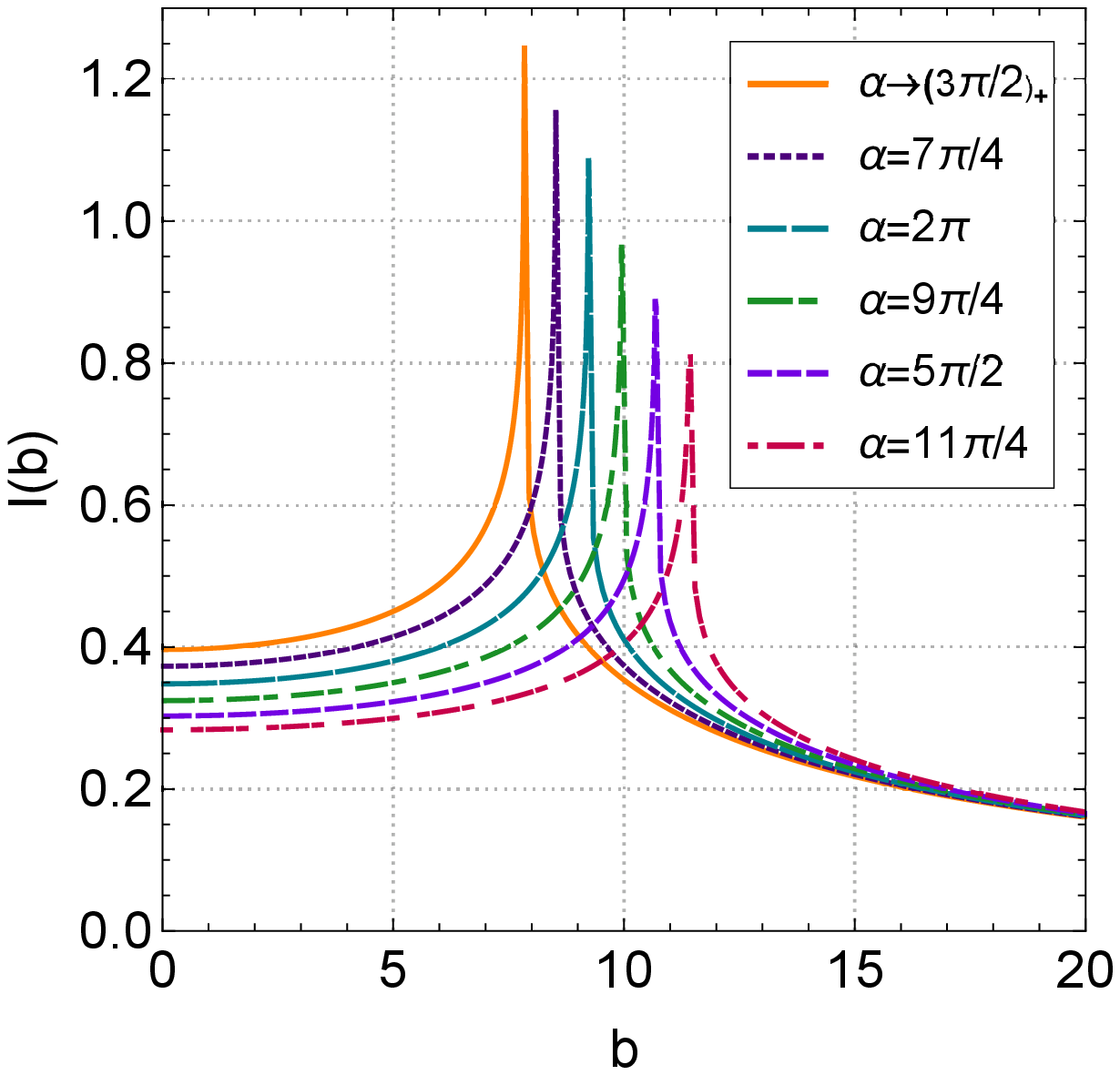}
\caption{The observed specific intensity at spatial infinity with different phantom parameter $\alpha$. The left and right panels correspond to $0<\alpha\leq\frac{3\pi{M}}{2}$ and $\alpha\geq\frac{3\pi{M}}{2}$, respectively. Here we set $M=1$. }
\label{as1}
\end{figure}
\begin{figure}
\includegraphics[width=16cm ]{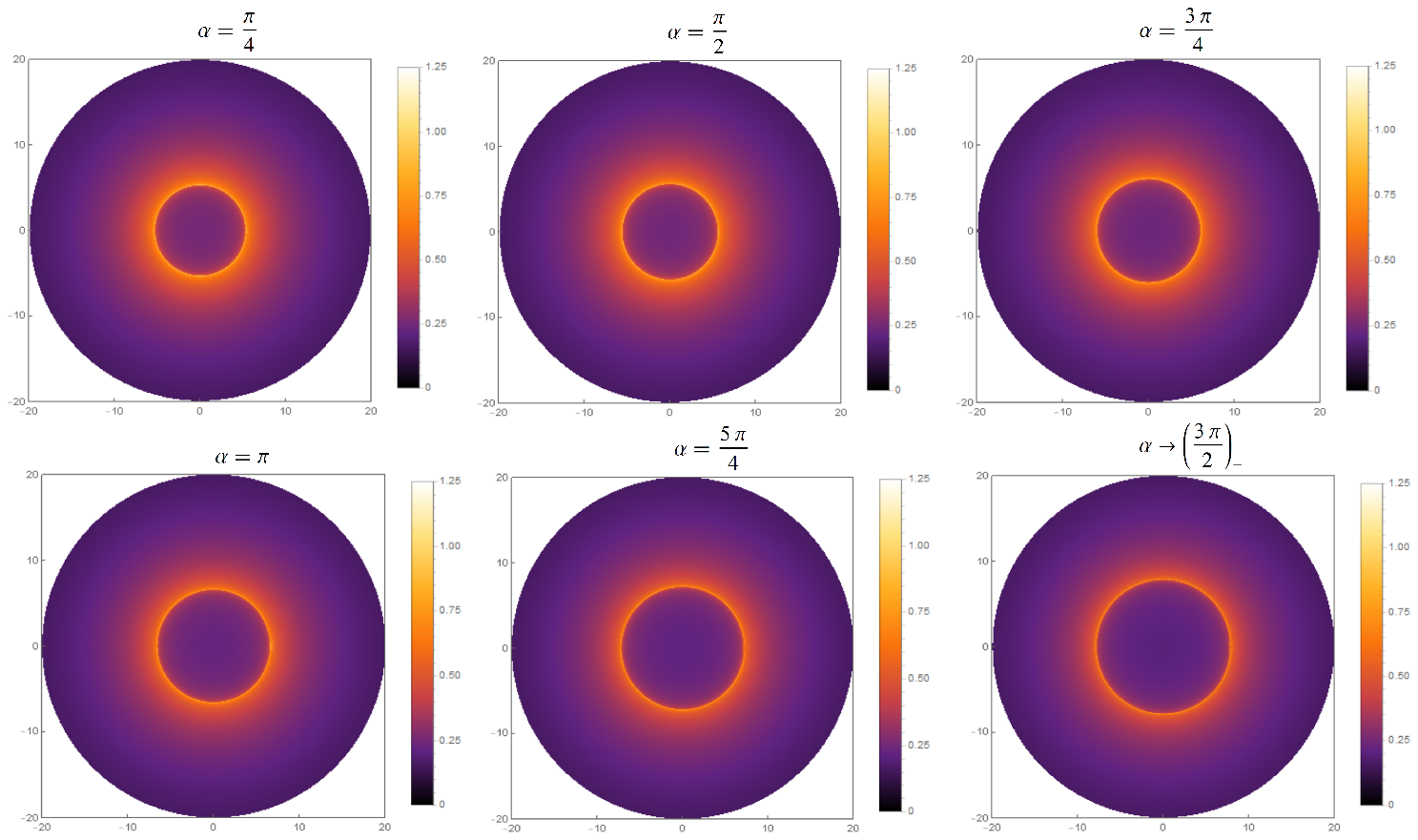}
\caption{Image of the regular phantom black hole under the static spherical accretion for different phantom parameter $\alpha$
 with $M=1$.}
\label{as2}
\end{figure}
\begin{figure}
\includegraphics[width=16cm ]{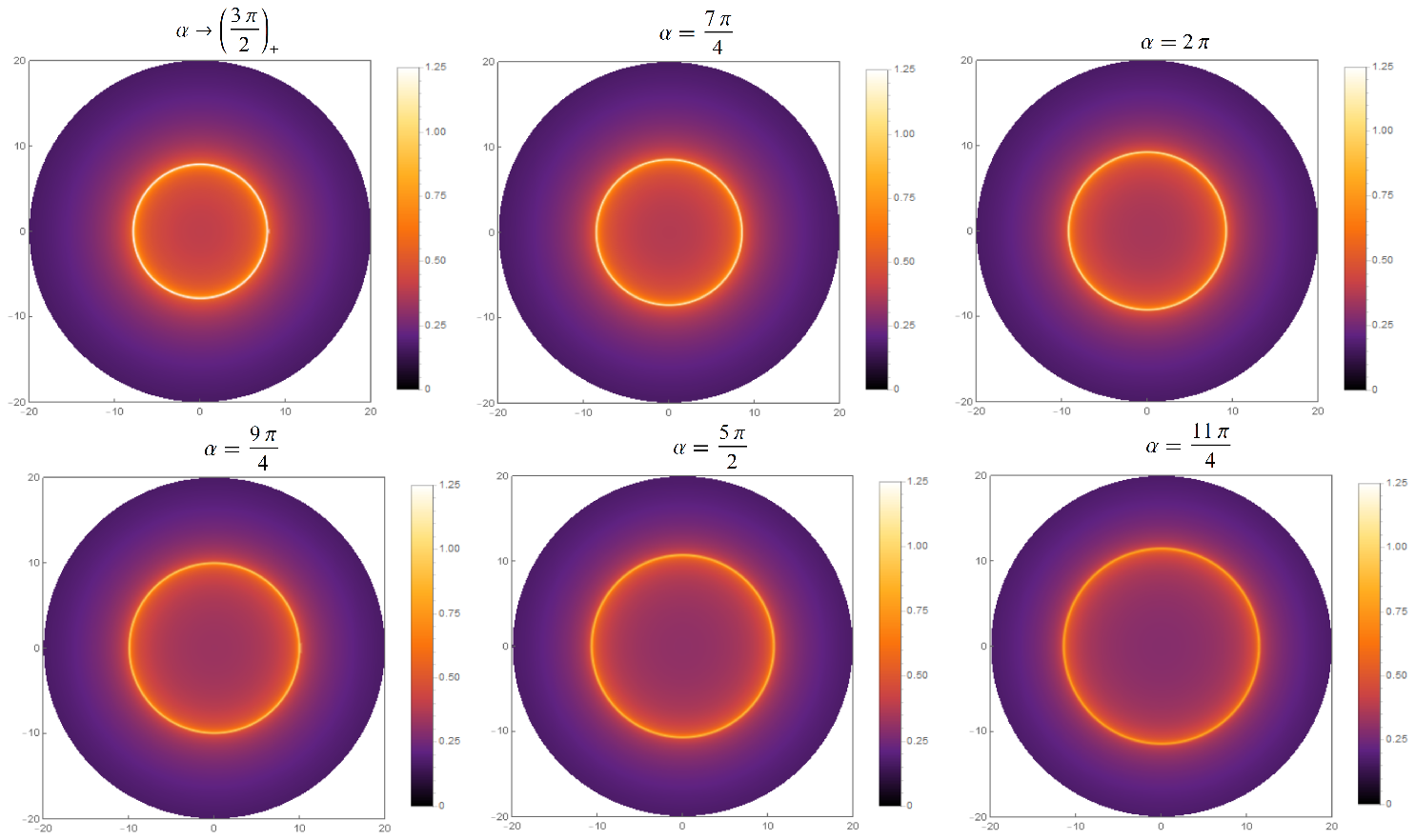}
\caption{Image of the regular phantom wormhole with the static spherical accretion for different phantom parameter $\alpha$
 with $M=1$.}
\label{as3}
\end{figure}
With Eq.(\ref{the radiation intensity}), we can analyze the shadow images and the corresponding intensities for the regular phantom compact object (\ref{newmetric}) using the rest gas model for a distant observer. In Fig.(\ref{as1}), we present the effects of phantom parameter $\alpha$ on the observed specific intensity at spatial infinity. As the compact object is a black hole, i.e., $\alpha<\frac{3\pi M}{2}$, one can find that with the increase of $b$ the specific intensity increases rapidly and reaches a peak at $b_{ph}$. However, with the further increasing $b$, it drops to a lower value. This behavior of the specific intensity with $b$ is similar to that in other black hole spacetimes, which could be a common feature for images of black holes.
With the increase of the phantom parameter $\alpha$, we find that the peak value of the specific intensity $I(b_{bh})$ decreases, which means that the maximum luminosity in the shadow images with the phantom scalar hair is less  than that of Schwarzschild black hole without phantom scalar hair. Moreover, we also obtain that the size of black hole shadow increase with the phantom parameter $\alpha$. From Fig.(\ref{as1}), one can find that the observed specific intensity $I(b)$ in the region of shadow (i.e., $b<b_{ph}$) decreases with the parameter $\alpha$, but in the region outside black hole shadow (i.e., $b>b_{ph}$) , it slightly increases with $\alpha$. This is different from that in the four-dimensional Gauss-Bonnet black hole where the observed specific intensity $I(b)$ always increases with the Gauss-Bonnet coupling parameter. As the phantom compact object is a wormhole, i.e., $\alpha>\frac{3\pi M}{2}$, we find the effects of phantom parameter $\alpha$ on the observed specific intensity at spatial infinity is similar to that in the case of phantom black hole. These properties of images of regular phantom compacted object and its luminosity are also shown in Fig.(\ref{as2}) and Fig.(\ref{as3}) where we present the shadow cast by the static spherical accretion mode in the $x,y$ plane for regular phantom black hole with different parameter $\alpha$. Moreover, from Figs.(\ref{as1}) and (\ref{lo1}), we find that there exists a sudden jump for the peak of the specific intensity as the compact object \eqref{newmetric} changes from black hole into wormhole.
This can be explained by the fact that for a wormhole,  photons after reaching the throat come back to observer at the spatial infinite rather than is absorbed by the compact object.

\subsection{Image of a regular phantom compact object with the infalling spherical accretion}

\begin{figure}
\includegraphics[width=6cm ]{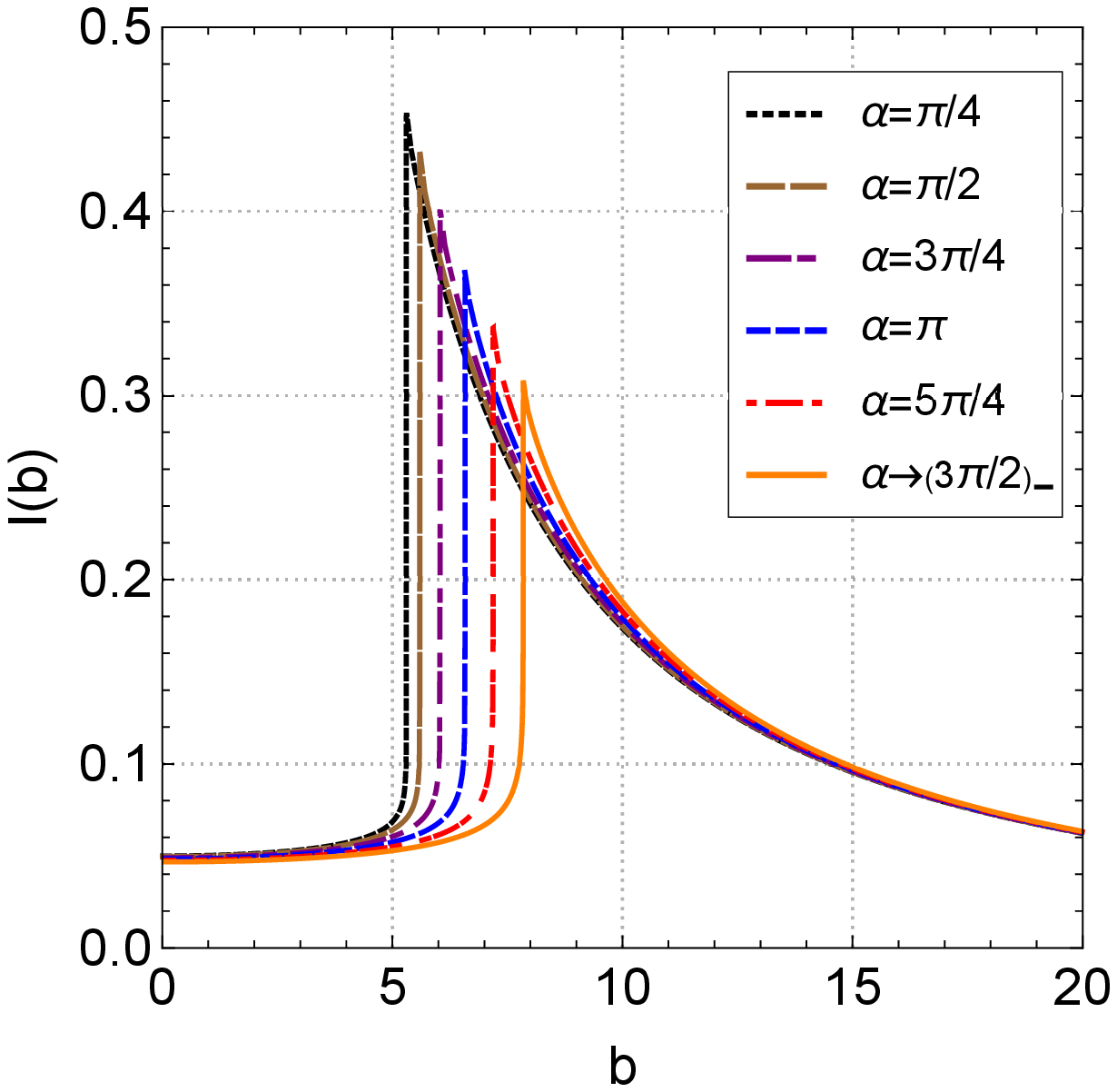}\quad\quad\includegraphics[width=6cm ]{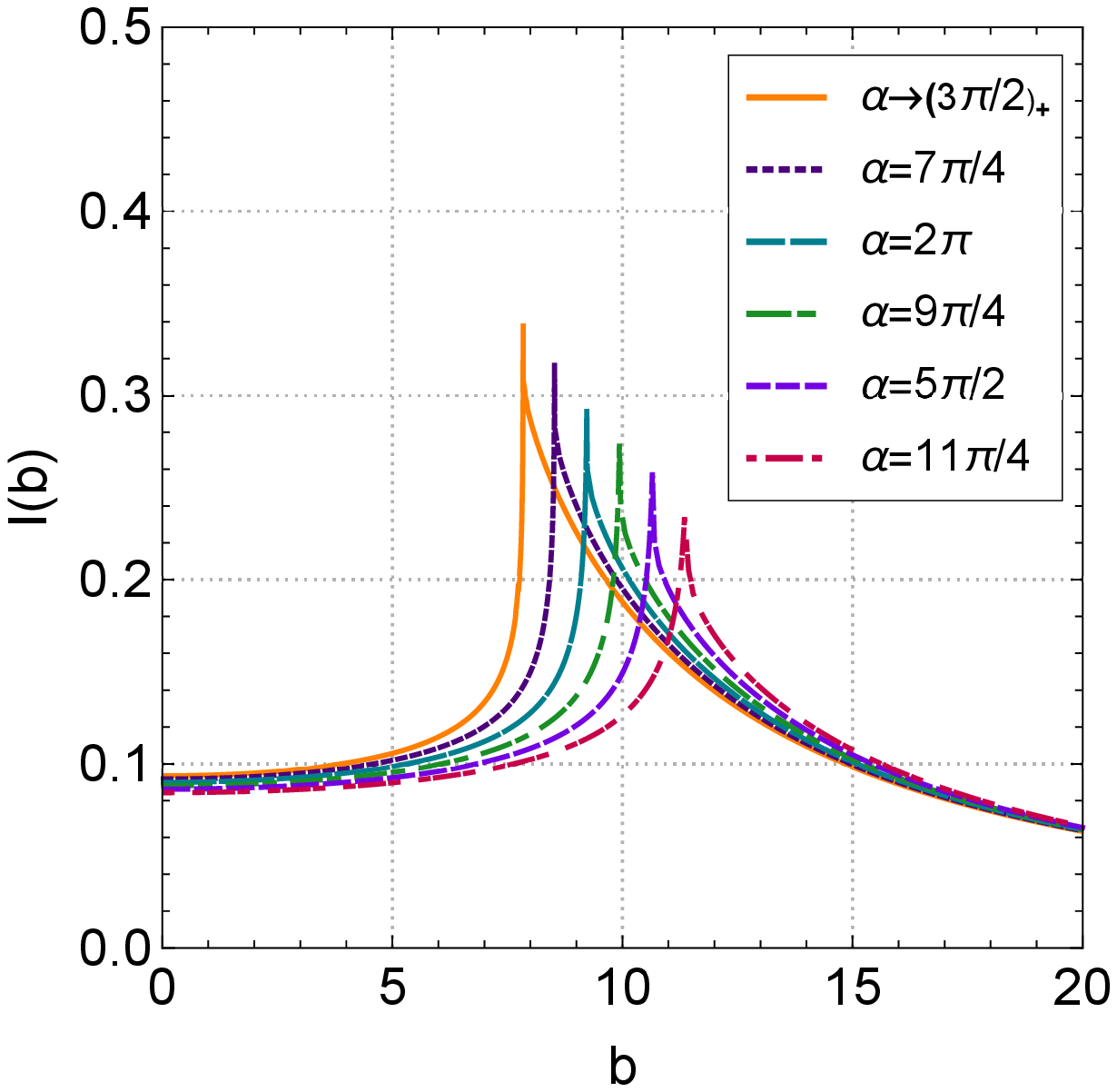}
\caption{The observed specific intensity  at spatial infinity with different phantom parameter $\alpha$ for the regular phantom black hole with the infalling spherical accretion. The left and right panels correspond to $0<\alpha\leq\frac{3\pi{M}}{2}$ and $\alpha\geq\frac{3\pi{M}}{2}$, respectively. Here we set $M=1$. }
\label{as4}
\end{figure}

\begin{figure}
\includegraphics[width=16cm]{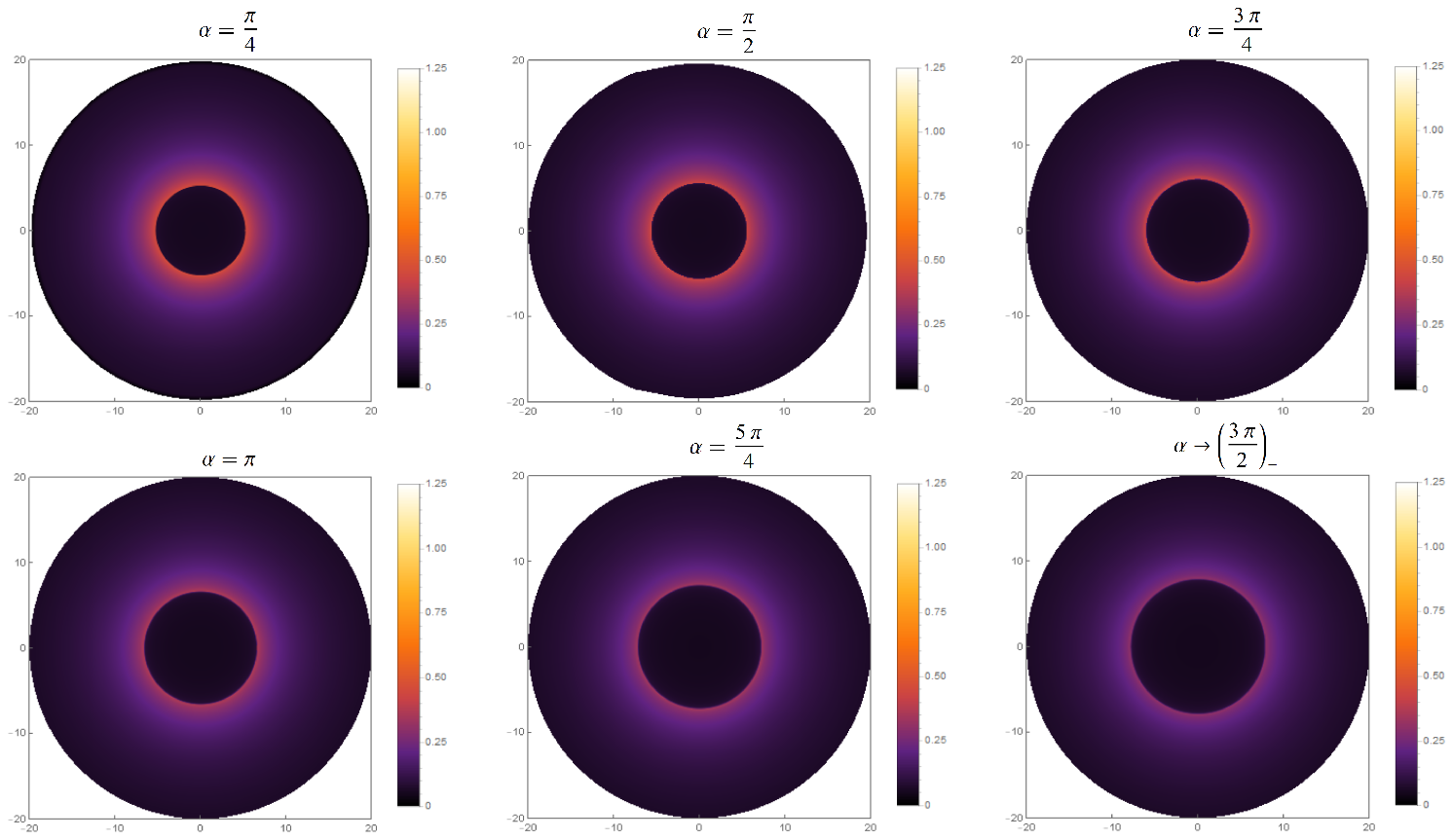}
\caption{Image of the regular phantom black hole under the infalling spherical accretion for different phantom parameter $\alpha$ with $M=1$.}
\label{as5}
\end{figure}

\begin{figure}
\includegraphics[width=16cm]{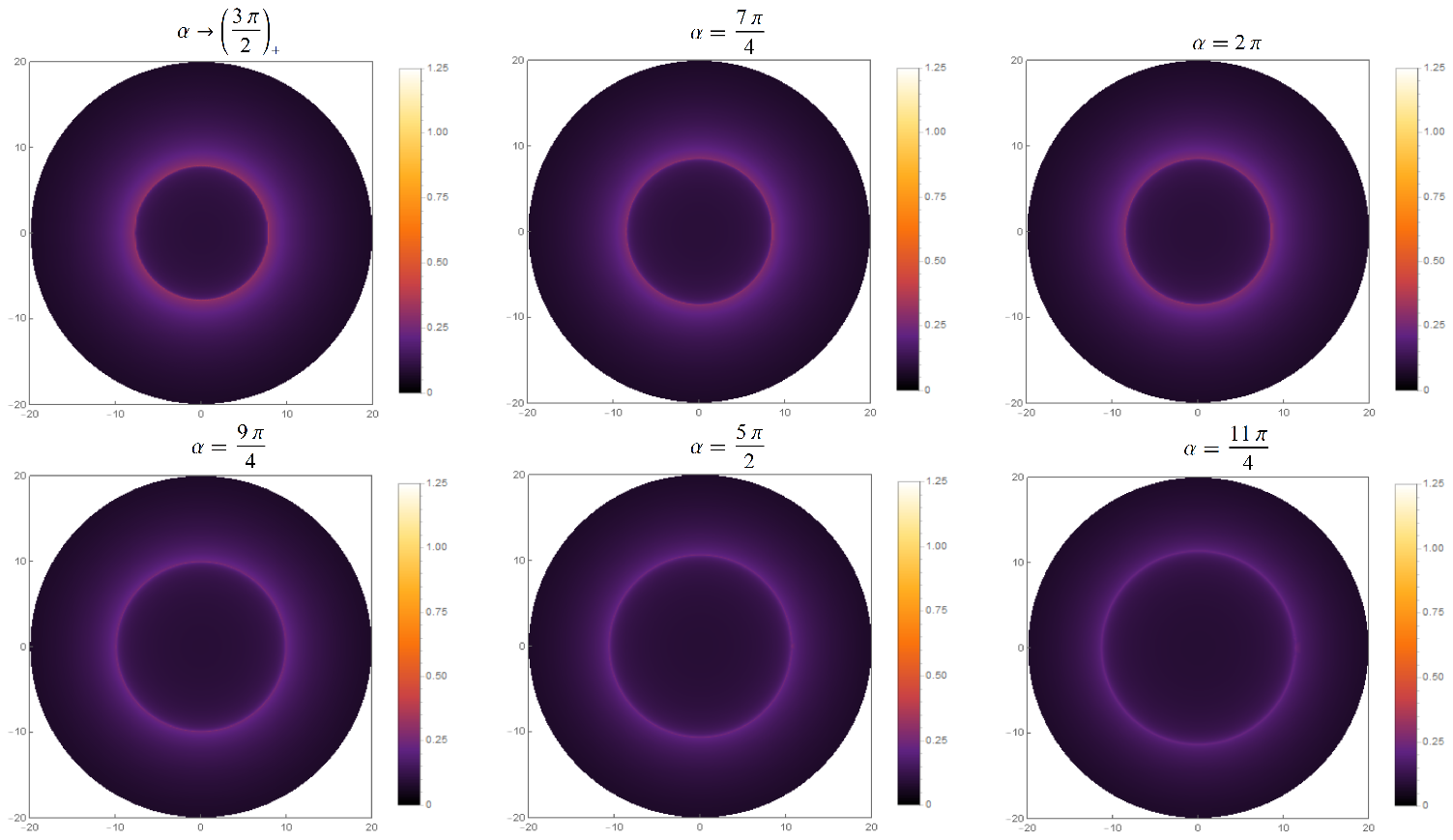}
\caption{Image of the regular phantom wormhole under the infalling spherical accretion for different phantom parameter $\alpha$ with $M=1$.}
\label{as6}
\end{figure}

Let us now to focus on the infalling spherical accretion around a regular phantom compact object where the radiating gas moves towards the compact object along the radial direction. In this case, due to the free falling motion of radiating gases, the corresponding redshift factor becomes
\begin{equation}\label{redshift factor2}
g=\frac{k_{\rho} u_o^{\rho}}{k_{\sigma} u_e^{\sigma}},
\end{equation}
where $k^\mu=\dot{x}^{\mu}$ is the four-velocity of the photon and $u_o^{\mu}=(1,0,0,0)$ is the four-velocity of the
distant observer. The quantity $u_e^{\mu}$ is the four-velocity of the accreting gas emitting the radiation, which has a form
\begin{equation}\label{four velocity accreting}
u_e^t=\frac{1}{F(r)},\quad\quad u_e^r=-\sqrt{\frac{r^2-\alpha^2}{r^2}(1-F(r))},\quad\quad u_e^{\theta}=u_e^{\phi}=0.
\end{equation}
From the null geodesic, we have the four-velocity of photons $k_\mu$ with a form
\begin{equation}\label{four velocity t}
k_t=\frac{1}{b},\quad\quad\quad
\frac{k_r}{k_t}=\pm\frac{1}{F(r)}\sqrt{\frac{r^2}{r^2-\alpha^2}\left(1-\frac{b^2F(r)}{r^2}\right)},
\end{equation}
where the sign $+$ or $-$ corresponds to the case that the photon approaches to or goes away from the compact object.
With these equations, the redshift factor (\ref{redshift factor2}) can be rewritten as
\begin{equation}
g=\frac{1}{u_e^{t}\pm\frac{k_{r}}{k_{t}}u_e^{r}}.
\end{equation}
The proper distance can be defined by \cite{NJG,xz,sau}
\begin{equation}
dl_{prop}=k_{\beta} u_e^{\beta}d\lambda=\frac{k_{t}}{g k^{r}}dr.
\end{equation}
Assuming the specific emissivity has the same form as in Eq.(\ref{emissivity}), the intensity $\eqref{intensity}$ in the case with infalling spherical accretion can be rewritten as
\begin{equation}\label{infalling intensity}
I(\nu_o)\varpropto\int_\gamma\frac{g^3k_tdr}{r^2k^r}.
\end{equation}
With above equation, we can study the image of the regular phantom compact object (\ref{newmetric}) and its luminosity using the infalling gas model observed at spatial infinity. The effects of the phantom parameter $\alpha$ on shadow image and the corresponding luminosity distribution are shown in Figs.(\ref{as4}), (\ref{as5}) and (\ref{as6}). We find that for each fixed phantom parameter, the intensity of the image in the infalling spherical accretion is darker than in the static accretion,  which could be attributed to the Doppler effect. The change of the maximum intensity appeared at photon ring and the observed specific intensity $I(b)$ with the phantom parameter $\alpha$ are similar to those in the case with static spherical accretion. As the compact object \eqref{newmetric} changes from black hole into wormhole, there also exists a sudden jump for the peak of the specific intensity, but the amplitude of jump is less than that in static accretion.
\begin{figure}
\includegraphics[width=6cm ]{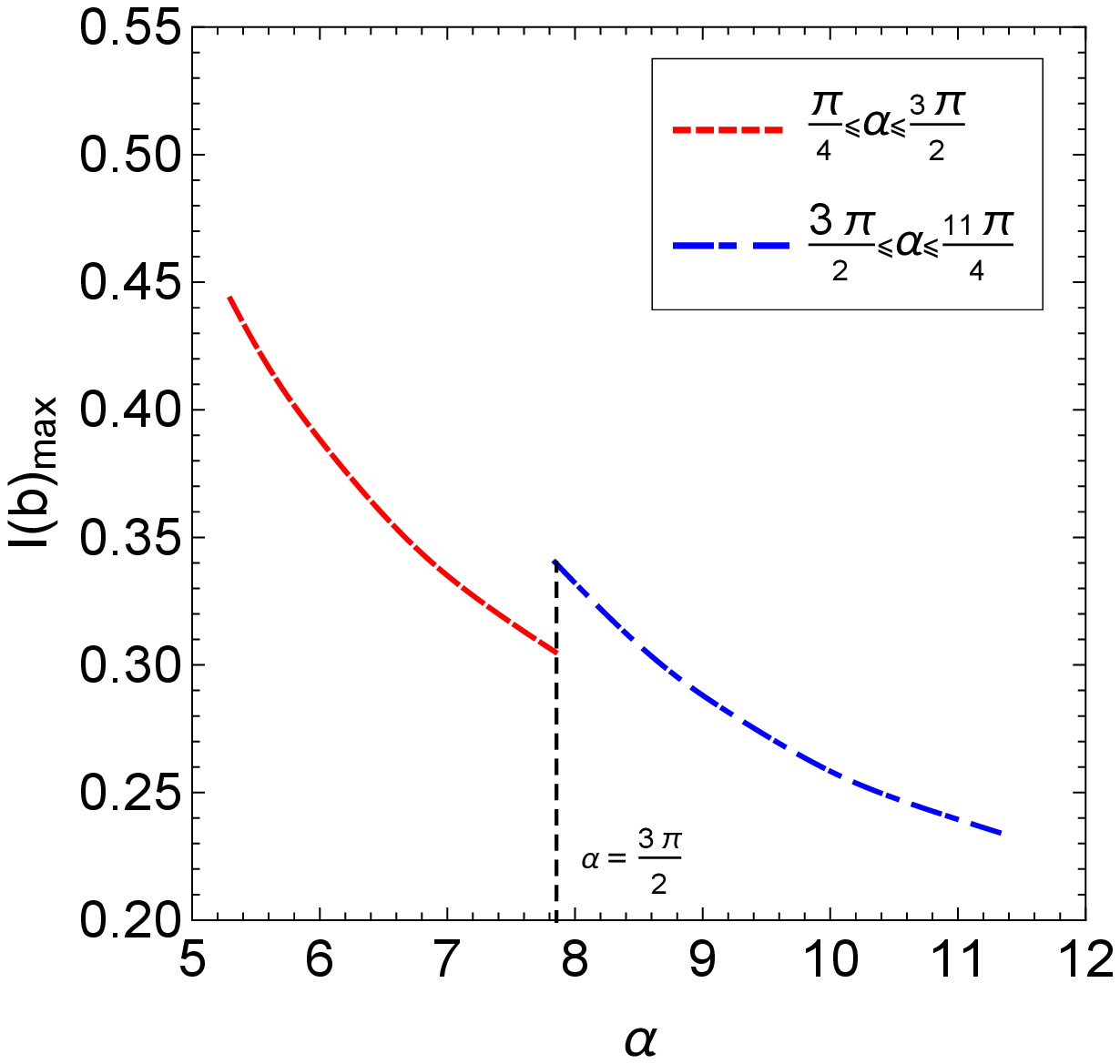}
\caption{The change of the the peak value of the specific intensity $I(b)$ with phantom parameter $\alpha$. Here we set $M=1$.}
\label{lo2}
\end{figure}

\section{The observed specific intensity of a regular slowly rotating phantom compact object }

Now, we consider the case of a slowly rotating regular phantom compact object. The metric can be expressed as \cite{phantom12}
\begin{equation}\label{ rotating metric}
ds^2=-F(r^\prime)dt^2+\frac{1}{F(r^\prime)}d{r^\prime}^2-2H(r^\prime,\theta)a{dtd\phi}+\left({r^\prime}^2+\alpha^2\right)
\left(d\theta^2+\sin^2{\theta}d\phi^2\right),
\end{equation}
with
\begin{equation}
H(r^\prime,\theta)=\frac{3M}{\alpha}\left[\left(\frac{\pi}{2}-\arctan\frac{r^\prime}{\alpha}\right)\left(1+\frac{{r^\prime}^2}{\alpha^2}\right)
-\frac{r^\prime}{\alpha}\right]\sin^2{\theta}.
\end{equation}
Where $a$ is the rotation parameter with its angular momentum. The event horizon of a slowly rotating phantom black hole is the same as that in the static and spherical symmetric case. As in the previous static case, we introduce a new coordinate $r^2={r^\prime}^2+\alpha^2$ and turn the metric form (\ref{ rotating metric}) into
\begin{equation}\label{new rotating metric}
ds^2=-F(r)dt^2+\frac{r^2}{F(r)\left(r^2-\alpha^2\right)}dr^2-2H(r,\theta)a{dtd\phi}+r^2\left(d\theta^2+\sin^2{\theta}d\phi^2\right),
\end{equation}
with
\begin{equation}
H(r,\theta)=\frac{3M}{\alpha}\left[\left(\frac{\pi}{2}-\arctan\frac{\sqrt{r^2-\alpha^2}}{\alpha}\right)
\frac{r^2}{\alpha^2}-\frac{\sqrt{r^2-\alpha^2}}{\alpha}\right]\sin^2{\theta}.
\end{equation}
For simplicity,  we here just consider that both the observer and the source lie in the
equatorial plane in the spacetime of a slowly rotating regular phantom compact object
(\ref{new rotating metric}) and the whole trajectory of the photon is limited on
the same plane. With this condition $\theta=\frac{\pi}{2}$, we find Lagrangian density for a particle can be obtained as
\begin{equation}\label{Rotating Lagrangian}
\mathcal{L}=\frac{1}{2}g_{\mu\nu}\dot{x}^\mu\dot{x}^\nu=\frac{1}{2}\left(-F(r)\dot{t}^2+
\frac{r^2}{F(r)\left(r^2-\alpha^2\right)}\dot{r}^2-2H(r)a\dot{t}\dot{\phi}+r^2\dot{\phi}^2\right),
\end{equation}
with
\begin{equation}
H(r)=\frac{3M}{\alpha}\left[\left(\frac{\pi}{2}-\arctan\frac{\sqrt{r^2-\alpha^2}}{\alpha}\right)
\frac{r^2}{\alpha^2}-\frac{\sqrt{r^2-\alpha^2}}{\alpha}\right].
\end{equation}
Adopting to the similar operation in the previous non-rotating case, we can obtain the timelike geodesic equations
\begin{equation}\label{timelike geodesics}
\dot{t}=\frac{r^2-bH(r)a}{br^2F(r)},\;\;\;\;\;\;\;\;\;\;\;
\dot{\phi}=\frac{bF(r)+H(r)a}{br^2F(r)},
\end{equation}
\begin{equation}
\frac{r^2}{\left(r^2-\alpha^2\right)}\dot{r}^2=-V_{eff}(r)=\frac{b^2F(r)+2abH(r)-r^2}{b^2r^2}.
\end{equation}
With the condition a circular orbit on the equator, $V_{eff}(r)=0$ and $\frac{dV_{eff}(r)}{dr}=0$, we can obtain the marginally circular orbit radius of the photon $r_{ph}$ and impact parameters $b_{ph}$. From Tab.(\ref{rotating taba}), one can see that for the fixed rotation parameter $a=0.3$,  both  $r_{ph}$ and $b_{ph}$  increase with the phantom charge $\alpha$, which is similar to those in the previous case of static black hole. From  Tab.(\ref{rotating tabaaa}), for the fixed phantom charge $\alpha=\pi/4$, both  $r_{ph}$ and $b_{ph}$ decrease with the rotation parameter $a$ as in the usual rotating black holes.
\begin{table}[H]
\centering
\caption{The marginally circular orbit radius of the photon $r_{ph}$ and the impact parameters $b_{ph}$. Here we set $M=1$ and $a=0.3$.}
\renewcommand\tabcolsep{18.0pt}
\begin{tabular*}{15cm}{llllllll}
\hline
&$\alpha$=0     & $\alpha$=$\frac{\pi}{4}$    & $\alpha$=$\frac{\pi}{2}$       & $\alpha$=$\frac{3\pi}{4}$        & $\alpha$=$\pi$        & $\alpha$=$\frac{5\pi}{4}$                 \\
\hline
\noalign{\global\arrayrulewidth 1 pt}\noalign{\global\arrayrulewidth 0.4 pt}
$r_{ph}$           &2.6007       & 2.7280                       & 3.0733                        & 3.5653                      & 4.1478                   & 4.7866                                                    \\
$b_{ph}$           &4.5012       & 4.6448                       & 5.0222                        & 5.5521                      & 6.1708                   & 6.8417                                             \\ \hline
\hline
 & $\alpha$=$\frac{3\pi}{2}$    & $\alpha$=$\frac{7\pi}{4}$       & $\alpha$=$2\pi$   & $\frac{9\pi}{4}$        & $\alpha$=$\frac{5\pi}{2}$      & $\alpha$=$\frac{11\pi}{4}$            \\
\hline
\noalign{\global\arrayrulewidth 1 pt}\noalign{\global\arrayrulewidth 0.4 pt}
$r_{ph}$           & 5.4619       & 6.1617                     & 6.8789                    & 7.6087                        & 8.3478                   & 9.0943                                              \\
$b_{ph}$           & 7.5446       & 8.2680                     & 9.0054                    & 9.7526                        & 10.5069                   & 11.2665                                               \\ \hline
\end{tabular*}\label{rotating taba}
\end{table}
\begin{table}[H]
\centering
\caption{The marginally circular orbit radius of the photon and the impact parameters $b_{ph}$. Here we set $M=1$ and $\alpha=\frac{\pi}{4}$.}
\renewcommand\tabcolsep{18.0pt}
\begin{tabular*}{17cm}{llllllll}
\hline
& $a$=$-0.3$    & $a$=$-0.2$     & $a$=$-0.1$     & $a$=$0$        & $a$=$0.1$      & $a$=$0.2$      & $a$=$0.3$      \\
\hline
\noalign{\global\arrayrulewidth 0.8 pt}\noalign{\global\arrayrulewidth 0.5 pt}
$r_{ph}$           & 3.4009        & 3.3066              & 3.2070             & 3.1011              & 2.9875            & 2.8641             & 2.7280                             \\
$b_{ph}$           & 5.8267        & 5.6615              & 5.4870             & 5.3010              & 5.1015            & 4.8846             & 4.6448                                \\ \hline
\end{tabular*}\label{rotating tabaaa}
\end{table}

\subsection{The light intensity of a regular slowly rotating phantom black hole with the static spherical accretion}
In this section, we will use equation (\ref{the radiation intensity}) to study the intensity of a regular slowly rotating phantom black hole with the static spherical accretion, where the proper length is
\begin{equation}
 dl_{prop}=\sqrt{\frac{r^2}{F(r)\left(r^2-\alpha^2\right)}+r^2\frac{d\phi^2}{dr^2}}dr^2,
\end{equation}
with
\begin{equation}
\frac{dr}{d\phi}=\pm\frac{bF(r)}{bF(r)+H(r)a}\sqrt{(r^2-\alpha^2)\bigg[\frac{r^2}{b^2}-F(r)-\frac{2aH(r)}{b}\bigg]}.
\end{equation}
In the case of a slowly rotating phantom black hole, the intensity expression can be further written as
\begin{equation}\label{rotating static intensity}
I=\int_\gamma\frac{F(r)^{\frac{3}{2}}}{r^2}\sqrt{\frac{r^2}{F(r)\left(r^2-\alpha^2\right)}+r^2\frac{d\phi^2}{dr^2}}dr^2
\end{equation}
\begin{figure}[ht]
\includegraphics[width=6cm ]{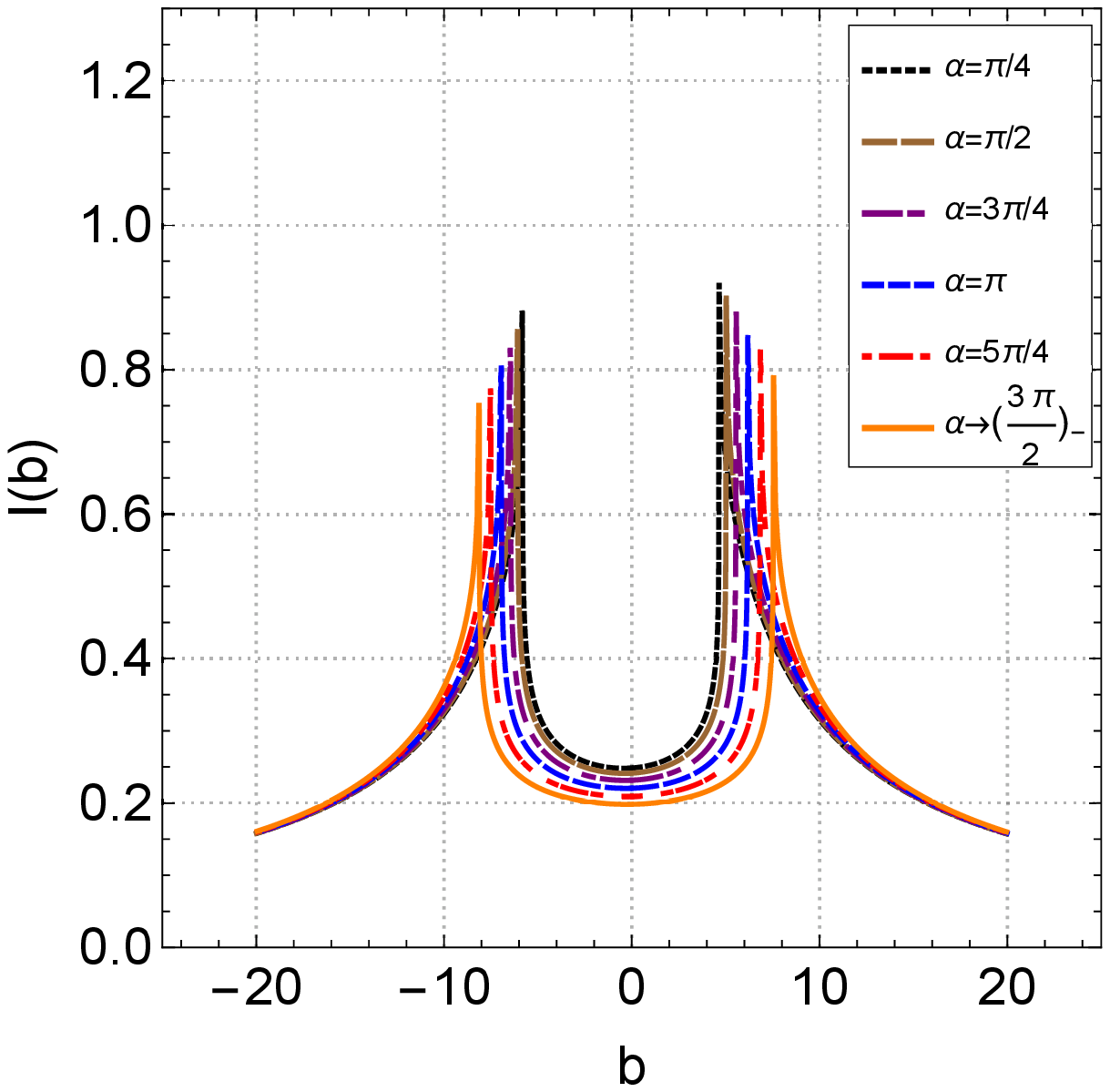}\quad\quad\includegraphics[width=6cm ]{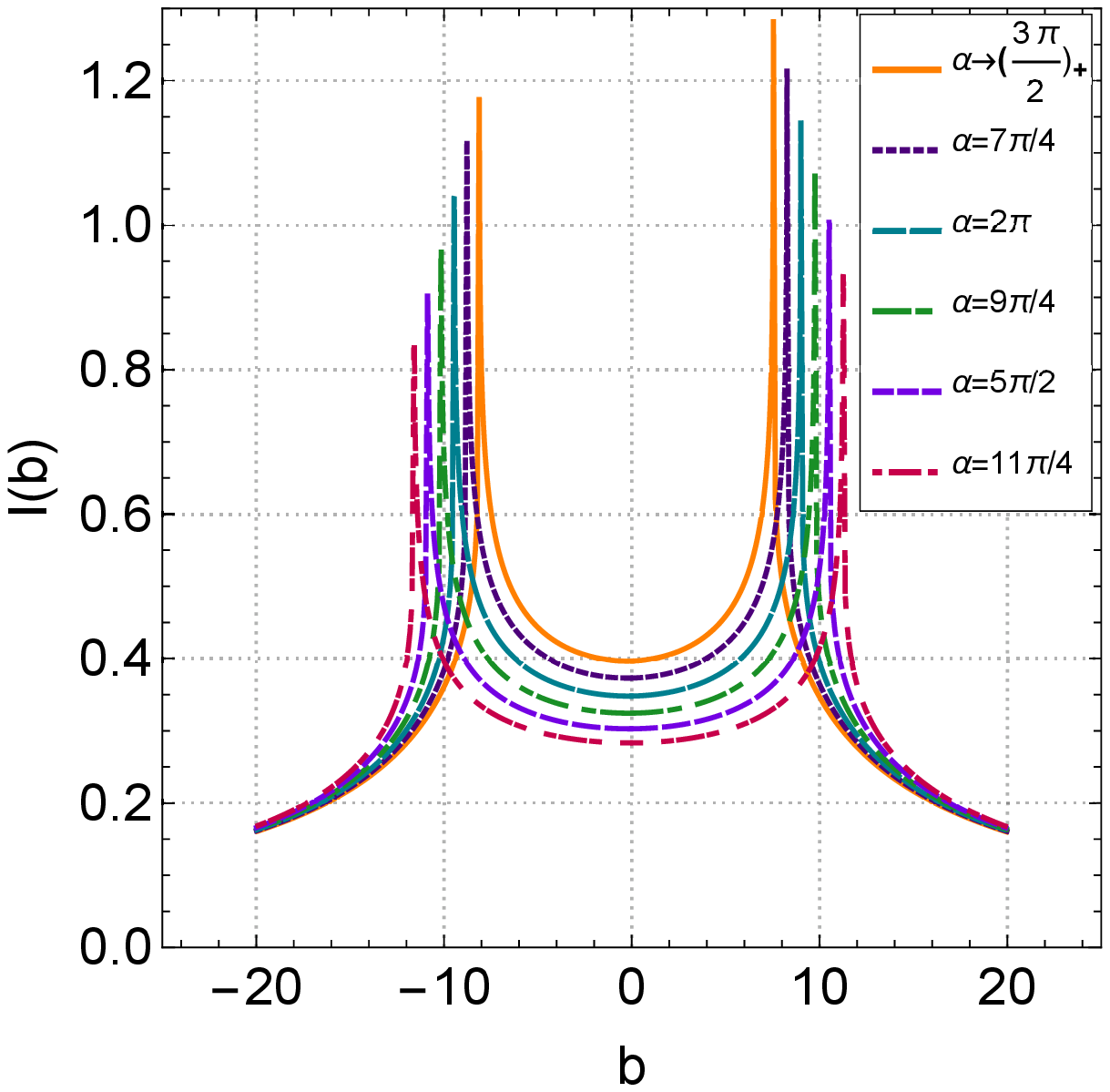}
\caption{The observed specific intensity at spatial infinity with different phantom parameter $\alpha$ for the regular slowly rotating phantom black hole with the static spherical accretion. The left and right panels correspond to $0<\alpha\leq\frac{3\pi{M}}{2}$ and $\alpha\geq\frac{3\pi{M}}{2}$, respectively. Here we set $M=1$, $a=0.3$. }
\label{rs1}
\end{figure}
From Fig.(\ref{rs1}), we can observe that in the both case of black hole and wormhole, for the fixed $\alpha$,  the peak intensity corresponding to the clockwise rotation  is higher than that to the counterclockwise rotation. Moreover, as $\alpha$ increases up to the critical value $\frac{3\pi}{2}$ at where the transition occurs between the black hole and the wormhole, there also exists a jump for the specific intensity.

\subsection{The light intensity of a regular slowly rotating phantom black hole with the infalling spherical accretion}

In a regular slowly rotating phantom black hole with the infalling spherical accretion, the four-velocity of the distant observer is $u_o^{\mu}=(1,0,0,0)$. The four-velocity of the photon can be expressed as
\begin{equation}
\kappa_t=F(r)\frac{r^2-bH(r)a}{br^2F(r)},
\end{equation}
\begin{equation}
 \frac{\kappa_r}{\kappa_t}=\pm\frac{1}{F(r)}\sqrt{\frac{r^2}{r^2-\alpha^2}\left[1-\frac{b^2F(r)}{r^2}
 -\frac{2ab^3F(r)^2}{r^4}\right]},
\end{equation}
Therefore, the redshift factor is
\begin{equation}
g=\frac{1}{\frac{1}{F(r)}\pm\frac{1}{F(r)}\sqrt{\bigg[1-\frac{b^2F(r)}{r^2}
 -\frac{2ab^3F(r)^2}{r^4}\bigg](1-F(r))}}.
\end{equation}
\begin{figure}
\includegraphics[width=6cm ]{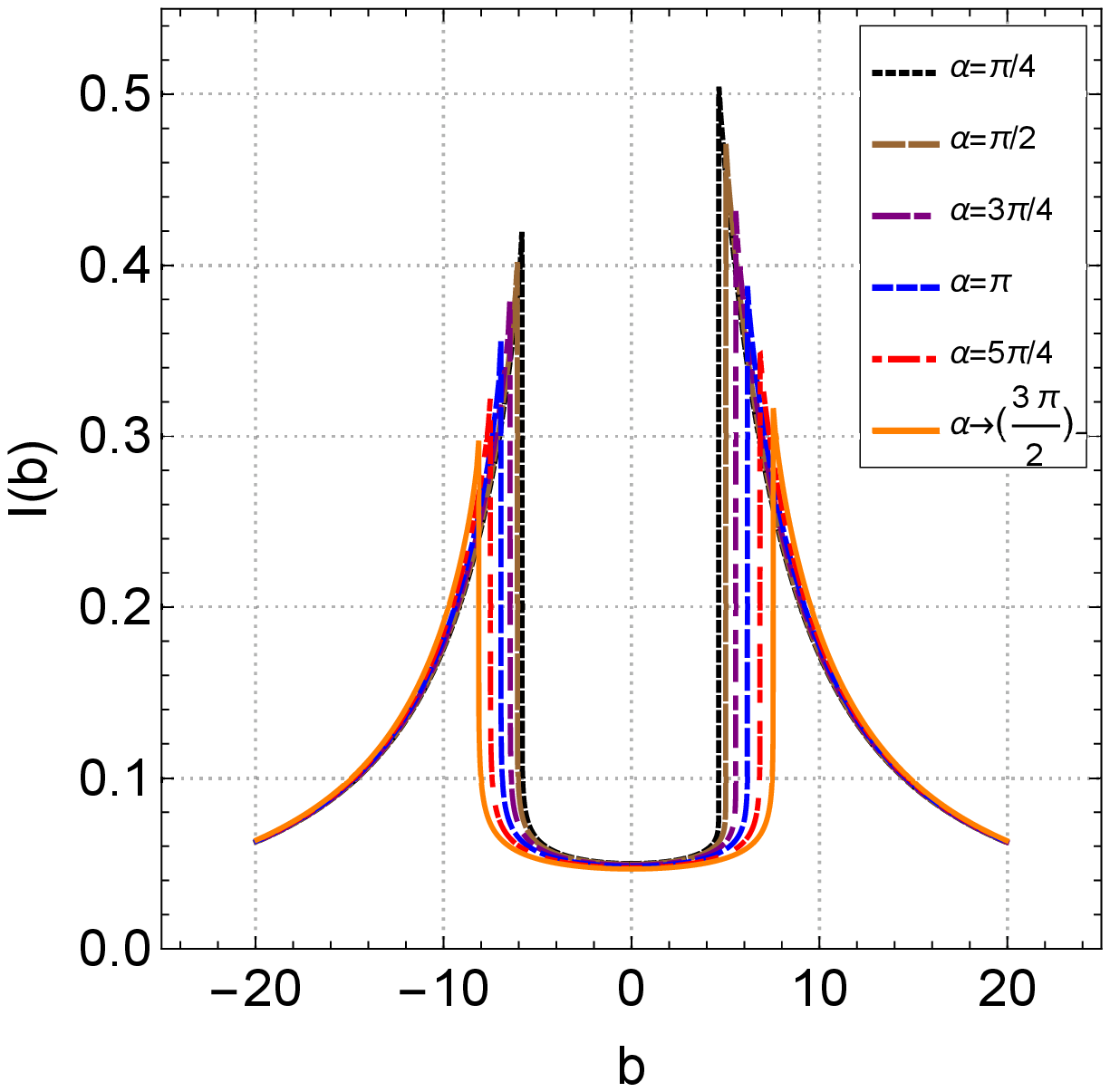}\quad\quad\includegraphics[width=6cm ]{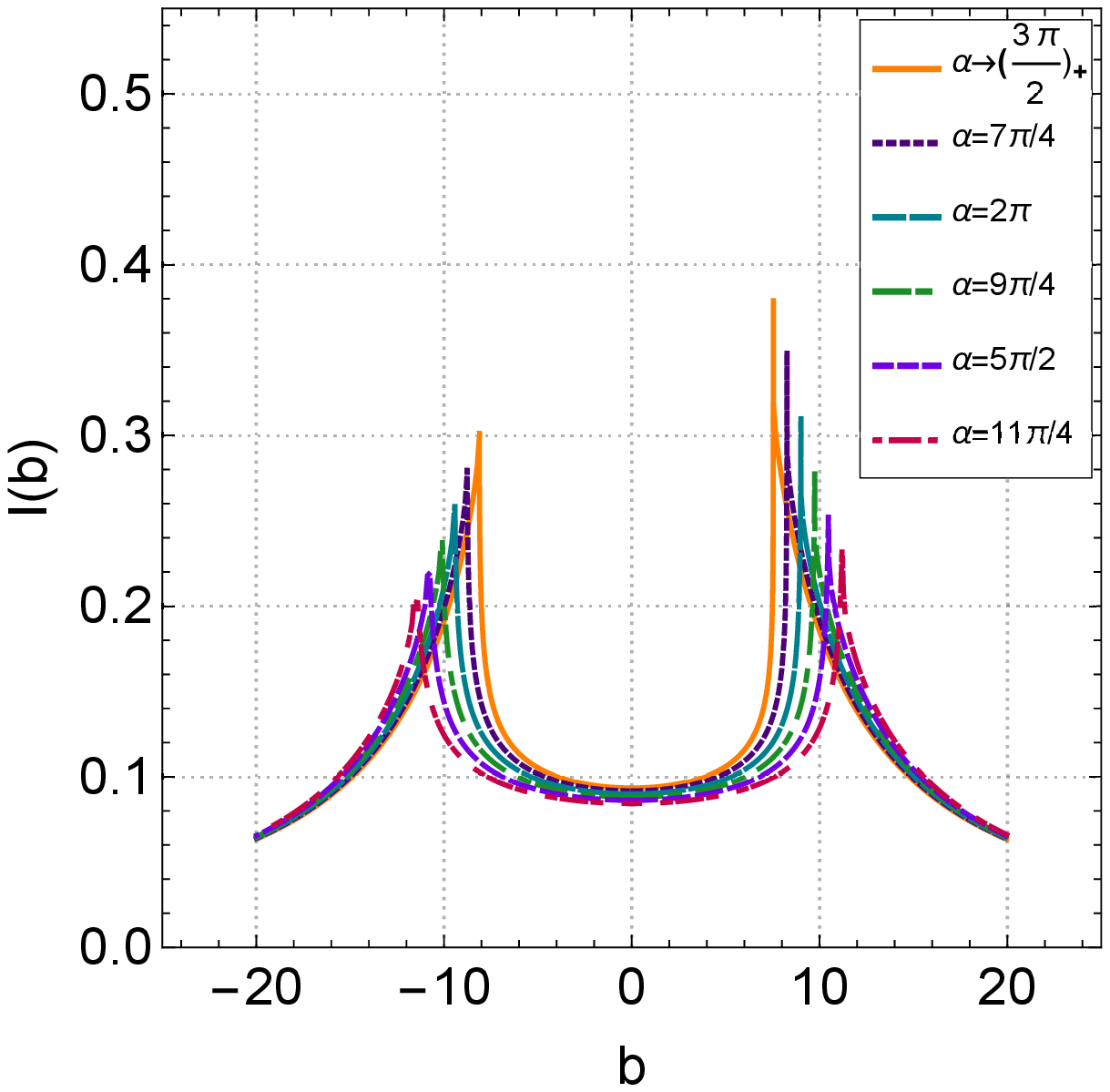}
\caption{The observed specific intensity at spatial infinity with different phantom parameter $\alpha$ for the regular slowly rotating phantom black hole with the infalling spherical accretion. The left and right panels correspond to $0<\alpha\leq\frac{3\pi{M}}{2}$ and $\alpha\geq\frac{3\pi{M}}{2}$, respectively. Here we set $M=1$, $a=0.3$. }
\label{ri1}
\end{figure}
Combining with Eq.(\ref{infalling intensity}), we finally present the specific intensity distribution in Fig.(\ref{ri1}). It is shown that as the phantom charge $\alpha$ increases, the shadow radius increases, but the peak intensity of the specific intensity decreases. Similarly, for the fixed $\alpha$,  the peak intensity in the infalling spherical accretion corresponding to the clockwise rotation is higher than that to the counterclockwise rotation, and a jump for the specific intensity also appears as the phantom charge parameter $\alpha$ increases up to the critical value $\alpha=\frac{3\pi}{2}$.

\section{Summary}

We have studied the image of a regular phantom compact object and its luminosity under the static spherical accretion and the infalling spherical accretion, respectively. It is shown that that both the photon sphere radius and the shadow size increases with the phantom parameter $\alpha$. In both spherical accretion models, with the increase of phantom parameter, the maximum luminosity occurred at photon ring and the brightness of the central region in the shadow decrease, but in the region far from the shadow, the luminosity of image slightly increases. This means that the presence of phantom hair yields the larger black hole shadow and the darker image by comparing with the usual Schwarzschild black hole. As the phantom charge parameter $\alpha$ increases up to the critical value $\frac{3\pi}{2}$ at where the transition occurs between the black hole and the wormhole, there exists a jump for the specific intensity. Moreover, in the case of infalling spherical accretion, the global brightness of the image becomes darker for each fixed phantom parameter, which is attributed to the known Doppler effect. The similar properties also appear in the case of a slowly rotating  phantom compact object.
These means that the phantom hair is imprinted on both the shadow radius and the intensity of the electromagnetic flux radiation around black hole.

\section{\bf Acknowledgments}

This work was  supported by the National Natural Science
Foundation of China under Grant No.11875026, 11875025, 12035005 and 2020YFC2201403.


\vspace*{0.2cm}

\begin{thebibliography}{99}
\baselineskip=0.5 cm
\bibitem{EHT1} The Event Horizon Telescope Collaboration, \textit{First M87 Event Horizon Telescope Results. I. The Shadow of the Supermassive Black Hole}, Astrophys. J. Lett. {\bf875}, L1 (2019).
\bibitem{EHT2} The Event Horizon Telescope Collaboration, \textit{First M87 Event Horizon Telescope Results. V. Physical origin of the asymmetric ring}, Astrophys. J. Lett. {\bf875}, L5 (2019).
\bibitem{EHT3} The Event Horizon Telescope Collaboration, \textit{First M87 Event Horizon Telescope Results. VI. The Shadow and Mass of the Central Black Hole}, Astrophys. J. Lett. {\bf875}, L6 (2019).


\bibitem{eht1} A. E. Broderick, T. Johannsen, A. Loeb and D. Psaltis, \textit{Testing the No-Hair Theorem with Event Horizon Telescope Observations of Sagittarius A* }, Astrophys. J. {\bf784}, 7 (2014).
\bibitem{nkerr} L. Medeiros, D. Psaltis, F. \"{O}zel, \textit{A Parametric Model for the Shapes of Black Hole Shadows in Non-Kerr Spacetimes}, Astrophys. J. {\bf896}, 7 (2020).

\bibitem{extr1} S. Vagnozzi and L. Visinelli, \textit{Hunting for extra dimensions in the shadow of M87*}, Phys. Rev. D {\bf100}, 024020 (2019).
\bibitem{extr2} I. Banerjee, S. Chakraborty, and S. SenGupta, \textit{Silhouette of M87*: A new window to peek into the world of hidden dimensions}, Phys. Rev. D {\bf 101}, 041301(R) (2020).

\bibitem{tomoch} Y. Chen, J. Shu, X. Xue, Q. Yuan, Y. Zhao, \textit{Probing Axions with Event Horizon Telescope Polarimetric Measurements}, Phys. Rev. Lett. {\bf124}, 061102 (2020).
\bibitem{dark1} R. Konoplya, \textit{Shadow of a black hole surrounded by dark matter}, Phys. Lett. B {\bf 795}, 1 (2019).
\bibitem{dark2}  X. Hou, Z. Xu, M. Zhou, J. Wang, \textit{Black Hole Shadow of Sgr A*   in Dark Matter Halo}, J. Cosmol. Astropart. Phys. {\bf1807}, 015 (2018).
\bibitem{dark3}  K. Jusufi, M. Jamil, P. Salucci, T. Zhu, S. Haroon, \textit{Black Hole Surrounded by a Dark Matter Halo in the M87 Galactic Center and its Identification with Shadow Images},  Phys. Rev. D {\bf100}, 044012 (2019).
\bibitem{Boson} H. Davoudiasl and P. B. Denton, \textit{Boson Dark Matter and Event Horizon Telescope Observations of M87*}, Phys. Rev. Lett. 123, no. 2, 021102 (2019)
\bibitem{dark4} P. V. P. Cunha, C. A. Herdeiro and E. Radu, \textit{EHT constraint on the ultralight scalar hair of the M87 supermassiveblack hole}, Universe {\bf5}, 220 (2019), [arXiv:1909.08039 [gr-qc]
\bibitem{epb}C. Li, S. Yan, L. Xue, X. Ren, Y. Cai, D. A. Easson, Y. Yuan, and H. Zhao, \textit{Testing the equivalence principle via the shadow of black holes}, Phys. Rev. Research {\bf2}, 023164 (2020), arXiv:1912.12629 [astro-ph].

\bibitem{sw} P. V. P. Cunha, C. Herdeiro, E. Radu and H. F. Runarsson, \textit{Shadows of Kerr black holes with scalar hair}, Phys. Rev. Lett. {\bf115}, 211102 (2015), arXiv:1509.00021;
\bibitem{swo} P. V. P. Cunha, C. Herdeiro, E. Radu and H. F. Runarsson, \textit{Shadows of Kerr black holes with and without scalar hair}, Int. J. Mod. Phys. D {\bf25}, 1641021 (2016), arXiv:1605.08293.
\bibitem{astro}F. H. Vincent, E. Gourgoulhon, C. Herdeiro and E. Radu, \textit{Astrophysical imaging of Kerr black holes with scalar hair}, Phys. Rev. D {\bf94}, 084045 (2016), arXiv:1606.04246.
\bibitem{chaotic} P. V. P. Cunha, J. Grover, C. Herdeiro, E. Radu, H. Runarsson, and A. Wittig, \textit{Chaotic lensing around boson stars and Kerr black holes with scalar hair}, Phys. Rev. D {\bf94}, 104023 (2016).


\bibitem{grr0} J. P. Luminet, \textit{Image of a Spherical Black Hole with Thin Accretion Disk}, Astron. Astrophy. {\bf 75}, 228 (1979).

\bibitem{grr} K. Wu, S. V. Fuerst, K. G. Lee, and G. Branduardiraymont, \textit{General Relativistic Radiative Transfer: Emission from Accreting Black Holes in AGN}, Chin. J. Astron. Astrophys {\bf6}, 205 (2006).
\bibitem{short} J. A. Marck, \textit{Short-cut method of solution of geodesic equations for Schwarzchild black hole}, Class. Quantum Grav. {\bf13}, 393 (1996).
\bibitem{BKD} K. Beckwith, C. Done, \textit{Extreme gravitational lensing near rotating black holes}, Mon. Not. R. Astron. Soc. {\bf359}, 1217 (2005).
\bibitem{gyoto} F. H. Vincent, T. Paumard, E. Gourgoulhon and G Perrin, \textit{GYOTO: a new general relativistic ray-tracing code}, Class. Quantum Grav. {\bf28}, 225011 (2011).

\bibitem{chensa} M. Wang, S. Chen, J. Wang, J. Jing, \textit{Shadow of a Schwarzschild black hole surrounded by a
Bach¨CWeyl ring}, Eur. Phys. J. C {\bf 110}, 80, (2020).
\bibitem{cunha}P. V. P. Cunha, N. A. Eir, C. A. R. Herdeiro and J. P. S. Lemos, \textit{Lensing and shadow
of a black hole surrounded by a heavy accretion disk}, J. Cosmol. Astropart. Phys. {\bf 2003}, 035 (2020).

\bibitem{Gralla} S. E. Gralla, D. E. Holz and R. M. Wald, \textit{Black Hole Shadows, Photon Rings, and
Lensing Rings}, Phys. Rev. D {\bf100}, 024018 (2019).
\bibitem{termed} H. Falcke, F. Melia and E. Agol, \textit{Viewing the shadow of the black hole at the galactic
center}, Astrophys. J. {\bf528}, L13 (2000).



 \bibitem{phantom} R. R. Caldwell, \textit{A Phantom Menace?} Phys. Lett. B {\bf545}, 23 (2002).

\bibitem{phantom1} A. Melchiorri, L. Mersini, C. J. Odman and M. Trodden, \textit{The state of the dark energy equation of state}, Phys. Rev. D {\bf68} 043509 (2003).



\bibitem{BKZ} K. A. Bronnikov and J. C. Fabris, \textit{Regular Phantom Black Holes}, Phys. Rev. Lett. {\bf96},251101 (2006).

\bibitem{phantom3} G. W. Gibbons, D. A. Rasheed, \textit{Dyson Pairs and Zero-Mass Black Holes}, Nucl. Phys. B {\bf476}, 515 (1996), arXiv:hep-th/9604177.
\bibitem{phantom4} G. Cl\'{e}ment, J. C. Fabris, M. E. Rodrigues, \textit{Phantom Black Holes in Einstein-Maxwell-Dilaton Theory}, Phys. Rev. D {\bf79}, 064021 (2009), arXiv:0901.4543.
\bibitem{phantom5} M. Azreg-Ainou, G. Cl\'{e}ment, J. C. Fabris, M. E. Rodrigues, \textit{Phantom black holes and sigma models}, Phys. Rev. D {\bf83}, 124001 (2011), arXiv:1102.4093.
\bibitem{phantom6} C. J. Gao, S. N. Zhang, \textit{Phantom Black Holes}, arXiv:hep-th/0604114
\bibitem{phantom7} M. E. Rodrigues, Z. A. A. Oporto, \textit{Thermodynamics of phantom black holes in Einstein-Maxwell-dilaton theory}, Phys. Rev. D {\bf85}, 104022 (2012), arXiv:1201.5337.
\bibitem{phantom8} D. F. Jardim, M. E. Rodrigues, M. J. S. Houndjo, Eur. Phys. J. Plus {\bf127}, 123 (2012), arXiv:1202.2830.
\bibitem{phantom9} A. Nakonieczna, M. Rogatko, R. Moderski, \textit{Dynamical collapse of charged scalar field in phantom gravity}, Phys. Rev. D {\bf86}, 044043 (2012), arXiv:1209.1203.
\bibitem{phantom10} S. V. Bolokhov, K. A. Bronnikov, M. V. Skvortsova, \textit{Magnetic black universes and wormholes with a phantom scalar}, Class. Quant. Grav. {\bf29}, 245006 (2012).
\bibitem{phantom11} S. Chen, J. Jing, \textit{Gravitational field of a slowly rotating black hole with
a phantom global monopole}, Class. Quantum Grav. {\bf30}, 175012 (2013).
\bibitem{phantom12} S. Chen, M. Wang, J. Jing, \textit{Testing gravity of a regular and slowly
rotating phantom black hole by quasiperiodic oscillations}, Class. Quantum Grav. {\bf33}, 195002 (2016).


\bibitem{lensing1} M. Azreg-Ainou, \textit{Light paths of normal and phantom Einstein-Maxwell-dilaton black holes}, Phys. Rev. D {\bf87}, 024012 (2013).
\bibitem{lensing2} C. Ding, C. Liu, Y. Xiao, L. Jiang, R.G. Cai, \textit{Strong gravitational lensing in a black-hole spacetime dominated by dark energy}, Phys. Rev. D {\bf88}, 104007 (2013).
\bibitem{lensing3} E. F. Eiroa, C.M. Sendra, \textit{Regular phantom black hole gravitational lensing}, Phys. Rev. D {\bf88}, 103007 (2013).
\bibitem{lensing4} G. N. Gyulchev, I. Z. Stefanov, \textit{Gravitational lensing by phantom black holes}, Phys. Rev. D {\bf87}, 063005 (2013).

\bibitem{phbhole1} Y. F. Cai, and D. A. Easson, \textit{Black holes in an asymptotically safe gravity theory with higher derivatives}, J. cosmol. Astropart. Phys. {\bf 1009}, 002 (2010).
\bibitem{phbhole2} Y. F. Cai, G. Cheng, J. Liu, M. Wang, and H. Zhang, \textit{Features and stability analysis of non-Schwarzschild black hole in quadratic gravity}, J. High Energ. Phys. {\bf 1601}, 108 (2016).


\bibitem{expression1} M. Jaroszynski and A. Kurpiewski, \textit{Near kerr black holes: spectra of advection dominated accretion flows}, Astron. Astrophys. {\bf326}, 419 (1997).

\bibitem{expression2} C. Bambi, \textit{Can the supermassive objects at the centers of galaxies be traversable wormholes? The first test of strong gravity for mm/sub-mm very long baseline interferometry facilities}, Phys. Rev. D {\bf87}, 107501 (2013).


\bibitem{NJG} R. Narayan, M. D. Johnson and C. F. Gammie, \textit{The Shadow of a Spherically Accreting
Black Hole}, Astrophys. J. {\bf885}, L33 (2019).

\bibitem{xz}X. Zeng, H. Zhang, H. Zhang, \textit{Shadows and photon spheres with spherical accretions
in the four-dimensional Gauss-Bonnet black hole}, Eur. Phys. J. C {\bf 80}, 872 (2020).
\bibitem{sau} Saurabh, K. Jusufi,  \textit{Imprints of Dark Matter on Black Hole Shadows using Spherical Accretions}, arXiv:2009.10599 [gr-qc].
\bibitem{xz1}X. Zeng,  H. Zhang, \textit{ Influence of quintessence dark energy on the shadow of black hole}, arXiv:2007.06333 [gr-qc].







\end{thebibliography}
\end{document}